\documentclass[fleqn]{llncs}

\usepackage{llncsdoc}
\usepackage{graphicx}
\usepackage{appendix}
\usepackage{amssymb}
\usepackage{dsfont}
\usepackage{amsmath}
\usepackage{cancel}
\usepackage{verbatim}
\usepackage{chngpage}

\begin{document}
\title{On Distributed Model Checking\\ of MSO on Graphs}
\author{St\'{e}phane Grumbach  \and Zhilin Wu}
\institute{INRIA-LIAMA\thanks{INRIA-LIAMA -- CASIA -- Chinese
Academy of Sciences -- PO Box 2728 -- Beijing 100080 -- PR China --
\email{Stephane.Grumbach@inria.fr,  zlwu@liama.ia.ac.cn}}}


\maketitle

\begin{abstract}
\vspace*{-5mm} We consider distributed model-checking of Monadic
Second-Order logic (MSO) on graphs which constitute the topology of
communication networks. The graph is thus both the structure being
checked and the system on which the distributed computation is
performed. We prove that MSO can be distributively model-checked
with only a constant number of messages sent over each link for
planar networks with bounded diameter, as well as for networks with
bounded degree and bounded tree-length. The distributed algorithms
rely on nontrivial transformations of linear time sequential
algorithms for tree decompositions of bounded tree-width graphs.
\end{abstract}

\vspace*{-10mm}

\section{Introduction}
Model checking is a vital technique to formally verify finite-state
systems. Compared with the other verification techniques, such as
theorem proving, model checking enjoys the virtue that the
verification process can be fully automated. Formally, the model
checking problem for a given logic $\mathcal{L}$ is defined as
follows: given a sentence $\varphi$ in $\mathcal{L}$ and a finite
structure $\mathcal{S}$, check whether $\mathcal{S}$ is a model of
$\varphi$, i.e. whether $\mathcal{S} \models \varphi$. Model
checking has been widely used in the verification of circuits,
protocols, and software \cite{CGP00}.

Monadic second-order logic (MSO) is a second-order logic in which
second-order variables are restricted to set variables. MSO is of
great importance in the model checking community. Over words and
trees, MSO has been shown to have the same expressive power as
finite automata \cite{Thom97}. The temporal logics widely used in
model checking, such as LTL, CTL, modal mu-calculus, etc. can all be
seen as the fragments of MSO \cite{Emer90}. Moreover, MSO has been
applied directly to verify systems in practice. A model checker for
MSO, called MONA, has been developed to verify regular properties of
finite state systems \cite{HJJ+96}.

On the other hand, MSO on graphs are also very expressive. Many
interesting graph properties, e.g. 3-colorability, connectivity,
planarity, Hamiltonicity, etc. can be expressed \cite{Courcelle08}.

It is known that Model checking for MSO is PSPACE-complete
\cite{Vardi82}. This fact is often phrased as ``The combined
complexity for MSO model checking is PSPACE-complete''. Combined
complexity refers to the complexity in both the sentence and the
structure. In addition, two complexity measures, so called data
complexity and expression complexity, were introduced to distinguish
the complexity in resp. the structure and the sentence. The data
complexity refers to the problem of deciding whether a given
structure satisfies a fixed sentence, and the expression complexity
refers to the problem of deciding whether a given sentence holds in
a fixed structure. In general, expression complexity of model
checking problem is relatively high, for instance, even for positive
primitive formulas, that is, existentially quantified conjunctions
of atomic formulas, the expression complexity of model checking
problem is still NP-hard \cite{Vardi82}. On the other hand, the data
complexity of the model checking problem is in PTIME in many cases,
e.g. model checking for LTL, first-order logic (FO), etc.
Nevertheless, for MSO, the situation is a bit different: although
the data complexity for the MSO model checking on words and trees
are in PTIME, that on graphs is still NP-hard, since many NP-hard
problems, e.g. 3-colorability, can be expressed easily by MSO
sentences.

To deal with the high data complexity of MSO on graphs, restrictions
on graph classes can be made. The first seminal result in this
direction is Courcelle's theorem which shows that MSO model checking
on classes of graphs with bounded tree width has linear time data
complexity \cite{Courcelle90}. Since it is a natural idea to use
graph logics, e.g. MSO, to specify properties of topology graphs of
networks, we might wonder whether we could get a counterpart of
Courcelle's theorem in distributed computing.

Declarative logical languages have been recently applied
successfully to distributed computing: the so-called declarative
networking approach used some rule-based logical language (a
distributed variant of DATALOG) to describe networking protocols
\cite{LooCGGHMRRS06}. Inspired by the declarative networking
approach, in this paper, we start considering the distributed
computation of classical logical languages, such as MSO, which
express the properties of topology graphs of the network.
Specifically, we consider the distributed model checking of MSO on
classes of networks with bounded tree-width, motivated by getting a
distributed counterpart of Courcelle's theorem.

We consider communication networks based on the message passing
model \cite{AttiyaW04}, where nodes exchange messages with their
neighbors. The MSO sentences to be model-checked concern the graph
which form the topology of the network, and whose knowledge is
distributed over the nodes, which are only aware of their $1$-hop
neighbors.

Our main idea is to transform the centralized model checking
algorithm into a distributed one that is as efficient as possible.
The centralized model checking algorithm for MSO on bounded tree
width graphs works as follows: First a tree decomposition $T$ of the
given graph $G$ is computed, then the tree decomposition is
transformed into a labeled tree $T^\prime$, an automaton
$\mathcal{A}$ is obtained from the MSO sentence, then $\mathcal{A}$
is ran over $T^\prime$ in a bottom-up way to check whether $G$
satisfies the MSO sentence or not.

The main challenge in the transformation is how to distributively
construct and store a tree decomposition so that the automaton
obtained from the MSO sentence, can be efficiently ran over the
labeled tree obtained from the tree decomposition in a bottom-up
way.

We only obtain some partial results in this direction. We show that
MSO distributed model checking on several special but meaningful
classes of networks with bounded tree width, can be done with only a
\emph{constant} number of messages of size $O(\log n)$ ($n$ is the
number of nodes in the network) sent over each link. Specifically,
we show that over asynchronous distributed systems, MSO can be
distributively model-checked on planar networks with bounded
diameter, with only $O(1)$ messages sent per link; and over
synchronous distributed systems or asynchronous systems with a
pre-computed Breath-first-search (BFS) tree, MSO can be
distributively model-checked on networks with bounded degree and
bounded tree-length, with $O(1)$ messages sent per link.

Classes of networks with bounded tree width is in general of
unbounded tree length, vertices in the same bag of a tree
decomposition of the network may be arbitrarily far away from each
other, which makes it quite difficult, if not impossible, to
distributively construct tree decompositions for general networks
with bounded tree width, while ensuring the complexity bound that
only constant number of messages are sent over each link.

The constant bound on the number of messages sent over each link
ensures that the computation is {\it frugal} in the sense that it
relies on a bounded amount of knowledge for bounded degree graphs.
This assumption can be seen as a weakening of the locality property
introduced by  Linial \cite{Linial92}. Naor and Stockmeyer
\cite{NaorS95} showed that there were non-trivial locally checkable
labelings that are locally computable, while on the other hand
lower-bounds have been exhibited, thus resulting in non-local
computability results \cite{KuhnMW04,KuhnMW06}. Although logical
languages cannot be model-checked locally, their potential for
frugal computation is of great interest.

The paper is organized as follows. Preliminaries are presented in
the next section. In Section~\ref{sec-lintime}, we recall the sketch
of the proof of the linear time complexity of MSO on graphs with
bounded tree width. In Section~\ref{sec-dist-comp}, we define the
distributed computational model and exemplify the distributed model
checking of MSO by considering tree networks. Then in
Section~\ref{sec-MSO-planar}, we consider planar networks with
bounded diameter. Finally in Section~\ref{sec-MSO-BTL}, we consider
networks with bounded degree and bounded tree-length.


\vspace*{-3mm}

\section{Graphs, tree decompositions and logics}

\vspace*{-3mm} In this paper, our interest is focused to a
restricted class of structures, namely finite graphs. Let $G=(V,E)$,
with $E\subset V\times V$, be a finite graph. We use the following
notations. If $v \in V$, then $deg(v)$ denotes the {\it degree} of
$v$. For two nodes $u,v \in V$, the {\it distance} between $u$ and
$v$, denoted $dist_G(u,v)$, is the length of the shortest path
between $u$ and $v$. For $k \in \mathds{N}$, the {\it
$k$-neighborhood} of a node $v$, denoted $N_k(v)$, is defined as
$\{w \in V | dist_G(v,w) \le k\}$. If $\bar{v}=v_1...v_p$ is a
collection of nodes in $V$, then the $k$-neighborhood of $\bar{v}$,
denoted $N_k(\bar{v})$, is defined by $\bigcup_{1 \le i \le p}
N_k(v_i)$.
For $T\subseteq V$, let $\langle T \rangle^G$ denote the subgraph
induced by $T$.

Let $G=(V,E)$ be a connected graph, a \emph{tree decomposition} of
$G$ is a rooted labeled tree $\mathcal{T}=(T,F, r, B)$, where $T$ is
the set of vertices of the tree, $F \subseteq T \times T$ is the
child-parent relation of the tree, $r \in T$ is the root of the
tree, and $B$ is a labeling function from $T$ to $2^V$, mapping
vertices $t$ of $T$ to sets $B(t)\subseteq V$, called \emph{bags},
such that
\vspace*{-2mm}
\begin{enumerate}
\item For each edge $(v,w) \in E$, there is a $t \in T$, such that $\{v,w\} \subseteq B(t)$.
\item For each $v \in V$, $B^{-1}(v)=\{t \in T | v \in B(t)\}$ is connected in
$T$.
\end{enumerate}
\vspace*{-2mm}
The {\it width} of $\mathcal{T}$, $width(\mathcal{T})$, is defined
as $\max\{|B(t)|-1 | t \in T\}$. The tree-width of $G$, denoted
$tw(G)$, is the minimum width over all tree decompositions of $G$.
An \emph{ordered tree decomposition} of width $k$ of a graph $G$ is
a rooted labeled tree $\mathcal{T}=(T,F,r,L)$ such that:
\vspace*{-2mm}
\begin{itemize}
\item $(T,F,r)$ is defined as above,

\item $L$ assigns each vertex $t \in T$ to a $(k+1)$-tuple
$\overline{b^t}=(b^t_1,\cdots,b^t_{k+1})$ of vertices of $G$ (note
that in the tuple $\overline{b^t}$, vertices of $G$ may occur
repeatedly),

\item If $L^\prime(t):=\{b^t_j| L(t)= (b^t_1,\cdots,b^t_{k+1}), 1 \le j \le k+1\}$, then
$(T,F,r,L^\prime)$ is a tree decomposition.
\end{itemize}
\vspace*{-2mm}
The \emph{rank} of an (ordered) tree decomposition is the rank of
the rooted tree, i.e. the maximal number of children of its
vertices.

We consider monadic second-order logic (MSO) over the signature $E$,
where $E$ is a binary relation symbol. MSO is obtained by adding set
variables, denoted with uppercase letters, and set quantifiers into
first-order logic, such as $\exists X \varphi(X)$ (where $X$ is a
set variable). The reader can refer \cite{EbbinghausFlum99} for the
detailed syntax and semantics of MSO. MSO has been widely studied in
the context of graphs for its expressive power. For instance,
colorability, transitive closure or connectivity can be defined in
MSO \cite{Courcelle08}.



%


\vspace*{-3mm}

\section{Linear time centralized model-checking}\label{sec-lintime}

\vspace*{-3mm}

In this section, we consider the centralized model-checking of MSO,
and recall the main steps of the proof that MSO model checking over
classes of bounded tree-width graphs has the linear time data
complexity \cite{Courcelle90,FlumG06,FFG02}.

Let $\Sigma$ be some alphabet. A \emph{tree language} over alphabet
$\Sigma$ is a set of rooted $\Sigma$-labeled binary trees. Let
$\varphi$ be an MSO sentence over the vocabulary $\{E_1,E_2\}\cup
\{P_c | c \in \Sigma\}$, ($E_1,E_2$ are respectively the left and
right children relations of the tree),
 the tree language accepted by
$\varphi$, $\mathcal{L}(\varphi)$, is the set of rooted
$\Sigma$-labeled trees satisfying $\varphi$.

Tree languages can also be recognized by tree automata. A
\emph{deterministic bottom-up tree automaton} $\mathcal{A}$ is a
quintuple $(Q,\Sigma, \delta, f_0,F)$, where $Q$ is the set of
states; $F \subseteq Q$ is the set of final states; $\Sigma$ is the
alphabet; and
\vspace*{-2mm}
\begin{itemize}
\item $\delta: (Q \cup Q \times Q) \times \Sigma \rightarrow Q$ is the transition function; and

\item $f_0: \Sigma \rightarrow Q$ is the initial-state assignment function.
\end{itemize}
\vspace*{-2mm}
A \emph{run} of tree automaton $\mathcal{A}\!=\!(Q,\Sigma,
\delta,f_0,F)$ over a rooted $\Sigma$-labeled binary tree
$\mathcal{T}\!=\!(T,F,r,L)$ produces a rooted $Q$-labeled tree
$\mathcal{T}^\prime\!=\!(T,F,r,L^\prime)$ such that
\vspace*{-2mm}
\begin{itemize}
\item If $t \in T$ is a leaf, then $L^\prime(t)=f_0(t)$;

\item Otherwise, if  $t \in T$ has one child $t^\prime$,
then $L^\prime(t)=\delta(L^\prime(t^\prime),L(t))$;

\item Otherwise, if $t \in T$ has two children $t_1,t_2$,
then $L^\prime(t)=\delta(L^\prime(t_1),L^\prime(t_2),L(t))$.
\end{itemize}
\vspace*{-2mm}
Note that for each deterministic bottom-up automaton $\mathcal{A}$
and rooted $\Sigma$-labeled binary tree $T$, there is exactly one
run of $\mathcal{A}$ over $T$.

The \emph{run} $\mathcal{T}^\prime=(T,F,r,L^\prime)$ of
$\mathcal{A}=(Q,\Sigma, \delta,f_0,F)$ over a rooted
$\Sigma$-labeled binary tree $\mathcal{T}=(T,F,r,L)$ is
\emph{accepting} if $L^\prime(r) \in F$.

A rooted $\Sigma$-labeled binary tree $\mathcal{T}=(T,F,r,L)$ is
{\it accepted} by a tree automaton $\mathcal{A}=(Q,\Sigma,
\delta,f_0,F)$ if the run of $\mathcal{A}$ over $\mathcal{T}$ is
accepting. The tree language {\it accepted} by $\mathcal{A}$,
$\mathcal{L}(\mathcal{A})$, is the set of rooted $\Sigma$-labeled
binary trees accepted by $\mathcal{A}$.

The next theorem shows that the two notions are equivalent.

\vspace*{-2mm}

\begin{theorem}\label{thm-MSO-automata}
\cite{TW68} Let $\Sigma$ be a finite alphabet. A tree language over
$\Sigma$ is accepted by a tree automaton iff it is defined by an MSO
sentence. Moreover, there are algorithms to construct an equivalent
tree automaton from a given MSO sentence and to construct an
equivalent MSO sentence from a given automaton.
\end{theorem}
\vspace*{-2mm}

The centralized linear time algorithm for evaluating an MSO sentence
$\varphi$ over a graph $G=(V,E)$ with tree-width bounded by $k$
works as follows:
\vspace*{-2mm}
\begin{description}
\item[Step 1] Construct an ordered tree decomposition $\mathcal{T}=(T,F,r,L)$ of $G$ of width $k$ and rank $\le 2$;

\item[Step 2] Transform $\mathcal{T}$ into a $\Sigma_k$-labeled binary tree $\mathcal{T}^\prime=(T,F,r,\lambda)$ for some finite alphabet $\Sigma_k$;

\item[Step 3] Construct an MSO sentence $\varphi^\ast$ over vocabulary
$\{E_1,E_2\} \cup \{P_c| c \in \Sigma_k\}$ from $\varphi$ (over
vocabulary $\{E\}$) such that $G \models \varphi$ iff
$\mathcal{T}^\prime \models \varphi^\ast$;

\item[Step 4] From $\varphi^\ast$, construct a bottom-up binary tree automaton $\mathcal{A}$,
and run $\mathcal{A}$ over $\mathcal{T}^\prime$ to decide whether
$\mathcal{T}^\prime$ is accepted by $\mathcal{A}$.
\end{description}
\vspace*{-2mm}
For Step 1, it has been shown that a tree decomposition of graphs
with bounded tree-width can be constructed in linear time
\cite{Bodlaender93}. It follows from Theorem~\ref{thm-MSO-automata}
that Step 4 is feasible. The detailed description of Step 2 and Step
3 is tedious. Since the details of them are not essential in this
paper, we omit the detailed description of them here, and put it in
the appendix.


\vspace*{-4mm}

\section{Distributed model checking of MSO}\label{sec-dist-comp}

\vspace*{-3mm}

In the sequel, we present distributed algorithms to model-check MSO
over classes of networks of bounded tree-width. These algorithms are
obtained by transforming the centralized linear time algorithm
presented in the previous section into distributed ones, which admit
low complexity bounds. The challenge lies in two aspects. First, an
ordered tree decomposition could be distributively constructed, with
only $O(1)$ messages sent over each link. Second, the constructed
tree decomposition should be distributively stored in a suitable
way, so that the tree automaton obtained from the MSO sentence, can
be ran over the rooted labeled tree transformed from the ordered
tree decomposition, in a bottom-up way, still with only $O(1)$
messages sent over each link.

We consider a message passing model of computation \cite{AttiyaW04},
based on a communication network whose topology is given by a graph
$G=(V,E)$ of diameter $\Delta$, where  $E$ denotes the set of
bidirectional \emph{communication links} between nodes. From now on,
we restrict our attention to  \emph{finite connected graphs}.

Unless specified explicitly, we assume in this paper that the
distributed system is \emph{asynchronous} and has no failure. The
nodes have a unique \emph {identifier} taken from $1,2,\cdots,n$,
where $n$ is the number of nodes. Each node has distinct local ports
for distinct links incident to it. The nodes have {\it states},
including final accepting or rejecting states.


Let $\mathcal{C}$ be a class of graphs, and $\varphi$ an MSO
sentence, then we say that $\varphi$ can be distributively
model-checked over $\mathcal{C}$ if there exists a distributed
algorithm such that for each network $G \in \mathcal{C}$ and any
requesting node in $G$, the computation of the distributed algorithm
on $G$ terminates with the requesting node in the accepting state if
and only if $G \models \varphi$.

For the complexity of the distributed computation, we consider two
measures: the distributed time (TIME) and the maximal number of
messages sent over any link during the computation (MSG/LINK) with
message size $O(\log n)$.

\medskip

\noindent Let us first consider the simple case of trees to
exemplify the distributed model checking of MSO. In the centralized
model-checking of MSO over trees, it is necessary to encode the
(unranked) trees into binary trees.
The distributed model-checking of MSO sentence $\varphi$ over tree
networks is then carried on as follows:
\vspace*{-2mm}
\begin{itemize}
\item Through local replacement of each node $v$ by the set of (virtual)
nodes $\{[v,i]| 1\le i \le deg(v)\}$, the network is first
transformed into a (virtual) binary tree, and an ordered tree
decomposition $\mathcal{T}$ of width $1$ and rank $\le 2$ is
obtained;

\item The tree decomposition $\mathcal{T}$ is transformed into a
$\Sigma_2$-labeled binary tree $\mathcal{T}^\prime$;

\item The requesting node constructs a tree automaton $\mathcal{A}$ from
$\varphi$, and broadcasts $\mathcal{A}$ to all the nodes in the
network;

\item Finally $\mathcal{A}$ is ran distributively over $\mathcal{T}^\prime$ in
a bottom-up way to decide whether $\mathcal{T}^\prime$ is accepted
by $\mathcal{A}$.
\end{itemize}
\vspace*{-2mm}
%

%
\vspace*{-3mm}
\begin{example}[Distributed tree decomposition of tree networks]
The tree network and the ports of nodes are shown in
Fig.\ref{fig-tree-local-repl}(a). A rooted binary tree is obtained
by local replacement (Fig.\ref{fig-tree-local-repl}(b)). The ordered
tree decomposition is in Fig.\ref{fig-tree-local-repl}(c), the bags
$(i,j)$ satisfy that either $i=j$ or $j$ is the parent of $i$ in the
(original) tree network.
\vspace*{-4mm}
\begin{figure}[ht]
\centering
  \includegraphics[width=0.75\textwidth]{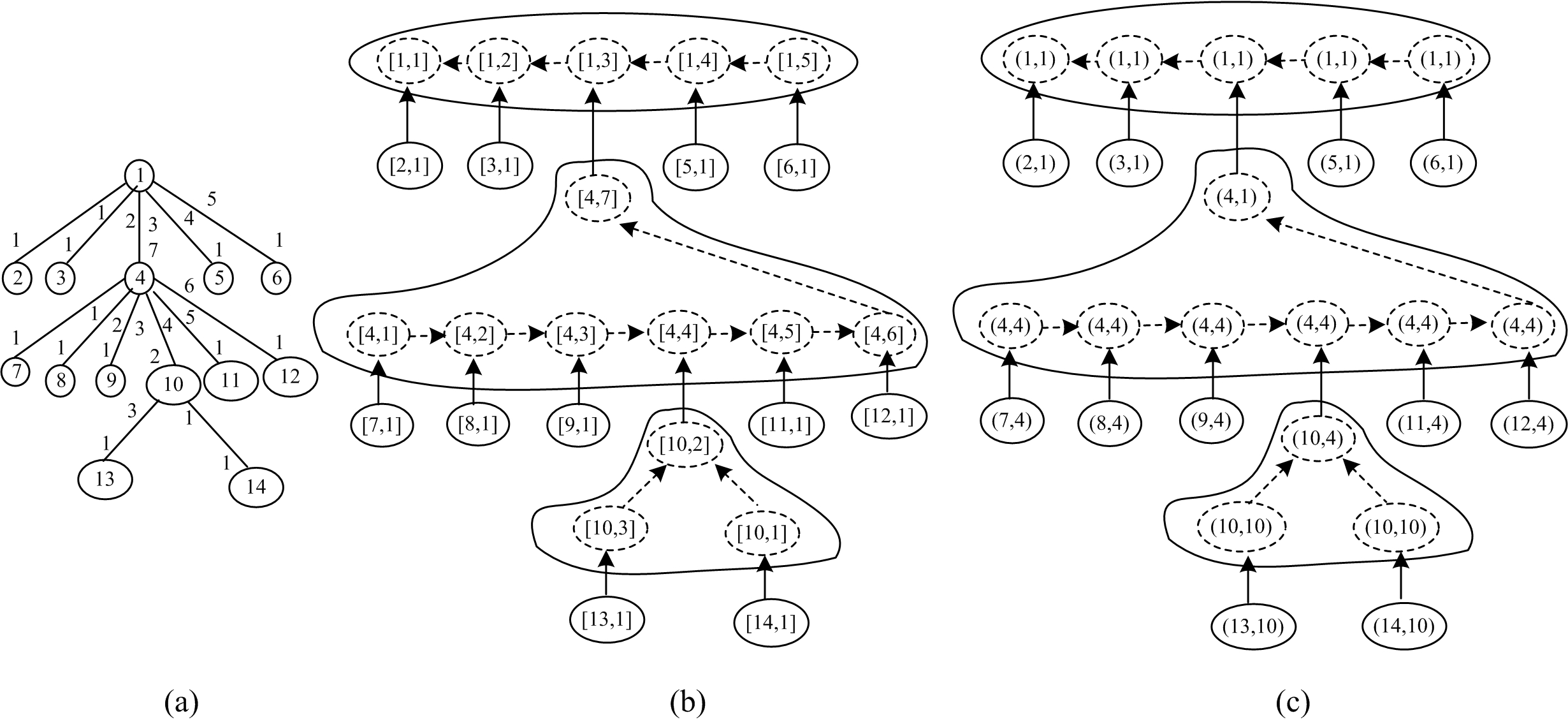}\\
  \caption{Distributed ordered tree-decomposition of tree network}\label{fig-tree-local-repl}
\end{figure}
\end{example}
\vspace*{-8mm}
%


\noindent Using the previous algorithm, we can prove the following.
\vspace*{-2mm}
\begin{theorem}
MSO can be distributively model-checked over tree networks within
complexity bounds $TIME=O(\Delta)$ and $MSG/LINK=O(1)$.
\end{theorem}
\vspace*{-2mm}

\vspace*{-3mm}

\section{Planar networks with bounded diameter}
\label{sec-MSO-planar}

\vspace*{-2mm}


We now consider planar networks with bounded diameter, and assume
that the diameter $k$ is known by each node. It has been shown that
if $G$ is a planar graph with diameter bounded by $k$, then the
tree-width of $G$ is bounded by $3k$ \cite{Eppstein95}.


A \emph{combinatorial embedding} of a planar graph $G=(V,E)$ is an
assignment of a cyclic ordering of the set of incident edges to each
vertex $v$ such that two edges $(u,v)$ and $(v,w)$ are in the same
face iff $(v,w)$ is put immediately before $(v,u)$ in the cyclic
ordering of $v$.
Combinatorial embeddings, which  encode the information about
boundaries of the faces in usual embeddings of planar graphs into
the planes, are useful for computing on planar graphs. Given a
combinatorial embedding, the boundaries of all the faces can be
discovered by traversing the edges according to the above condition.

\vspace*{-2mm}
\begin{example}[Combinatorial embedding] The combinatorial embedding and its corresponding usual embedding
into the planes are given in resp.
Fig.\ref{fig-combina-embedding-planar}(a) and
Fig.\ref{fig-combina-embedding-planar}(b). Suppose the edge
$\{a,d\}$ is traversed from $a$ to $d$, then the edge traversed next
is $\{d,b\}$, since in the cyclic ordering of $d$, $(d,b)$ is
immediately before $(d,a)$. Similarly, the edge traversed after
$\{d,b\}$ is $\{b,a\}$.

\vspace*{-6mm}

\begin{figure}[ht]
\centering
  \includegraphics[width=0.7\textwidth]{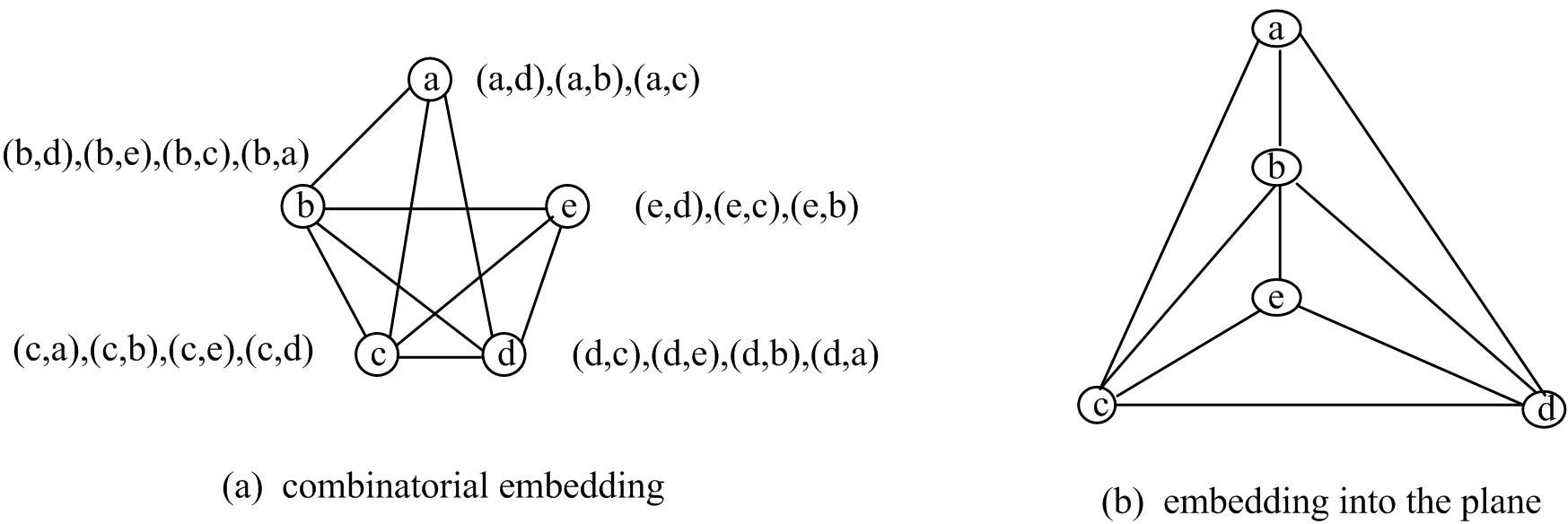}\\
  \caption{Combinatorial embedding and the corresponding usual
embedding into the planes}\label{fig-combina-embedding-planar}
\end{figure}
\end{example}

\vspace*{-8mm}

We assume in this section that a combinatorial embedding of the
planar network is distributively stored in the network, i.e. a
cyclic ordering of the set of the incident links is stored in each
node of the network.
\vspace*{-1.5mm}
\begin{theorem}\label{thm-MSO-sentence-planar-diameter}
MSO can be distributively model-checked over planar networks with
bounded diameter in complexity bounds $TIME=O(n)$ and $MSG/LINK =
O(1)$.
\end{theorem}
\vspace*{-1.5mm}
%

The main challenge of Theorem~\ref{thm-MSO-sentence-planar-diameter}
is the distributed construction and storage of an ordered tree
decomposition. In the following, we explain how to construct
distributively an ordered tree decomposition of width $3k$ for a
planar network with diameter bounded by $k$ such that the bags of
the tree decomposition are stored distributively in the nodes of the
network, and for each bag stored in $v$, its parent bag is stored in
some neighbor of $v$. If such an ordered tree decomposition has been
constructed, it can be transformed into a rooted
$\Sigma_{3k}$-labeled binary tree $\mathcal{T}$; the requesting node
then transforms the MSO sentence into a bottom-up tree automaton
$\mathcal{A}$ and broadcasts it to all the nodes in the network; and
$\mathcal{A}$ can be ran distributively over $\mathcal{T}$ in a
bottom-up way to check whether $\mathcal{A}$ accepts $\mathcal{T}$
by sending only $O(1)$ messages over each link.

We distinguish whether the planar network is biconnected or not.
\vspace*{-4mm}
\subsection{Biconnected planar networks with bounded diameter}
\vspace*{-3mm}
In this subsection, we assume that the planar networks are
biconnected. It is not hard to verify the following lemma.
\vspace*{-1.5mm}
\begin{lemma}
If a planar graph $G$ is biconnected, then the boundaries of all the
faces of a combinatorial embedding of $G$ are cycles.
\end{lemma}
\vspace*{-1.5mm}
These cycles are called the \emph{facial cycles} of the
combinatorial embedding.

We first recall the centralized construction of a tree decomposition
of a biconnected planar graph with bounded diameter
\cite{Eppstein95}.
\vspace*{-2mm}
\begin{itemize}
\item At first, the biconnected planar graph $G=(V,E)$ is triangulated into a planar graph $G^\prime$ by adding edges such that the boundary of each
face of $G^\prime$, including the outer face, is a triangle;

\item A breath-first-search (BFS) tree $T$ of $G^\prime$ is constructed;

\item An (undirected non-rooted) $2^V$-labeled tree $T^\prime=(I, F, L)$ is constructed such that

\begin{itemize}
\item $I$ is the set of faces of $G^\prime$;

\item $(i_1,i_2) \in F$ iff the face $i_1$ and $i_2$ have a common edge not
in the BFS-tree $T$;

\item $L(i)=ancestor(u) \cup ancestor(v) \cup ancestor(w)$, where $\{u,v,w\}$ are exactly the set of vertices contained in face $i$, where $ancestor(x)$ denotes the set of ancestors of $x$ in $T$;
\end{itemize}

\item Finally the tree decomposition is obtained from $T^\prime$ by selecting some vertex of $T^\prime$, i.e. face of $G^\prime$, to which the root of $T$ belongs,
as the root, and give directions to the edges of $T^\prime$
according to the selected root.
\end{itemize}
\vspace*{-5mm}
\begin{example}
[Ordered tree decomposition of biconnected planar graphs with
bounded diameter] A biconnected planar graph $G$ is given in
Fig.\ref{fig-centr-biconn-planar-diameter}(a). The triangulated
graph $G^\prime$ is in
Fig.\ref{fig-centr-biconn-planar-diameter}(b), with the dashed lines
denoting the edges added during the triangulation. A BFS tree of
$G^\prime$, $T$, is in
Fig.\ref{fig-centr-biconn-planar-diameter}(c), and a constructed
ordered tree decomposition $T^\prime$ is illustrated in
Fig.\ref{fig-centr-biconn-planar-diameter}(d), with the filled
circles denoting the faces (triangles) and arrows between them
denoting child-parent relationships. \qed
\vspace*{-6mm}
\begin{figure}[ht]
\centering
  \includegraphics[width=\textwidth]{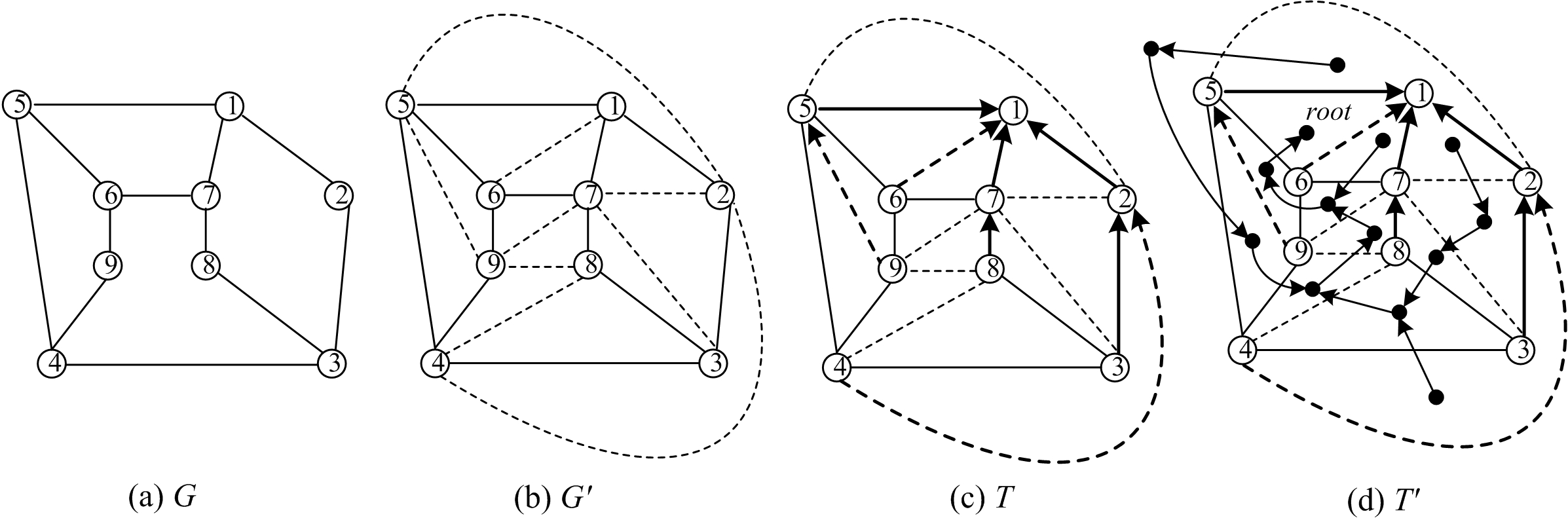}\\
  \caption{Centralized ordered tree decomposition of biconnected planar graphs with bounded diameter}\label{fig-centr-biconn-planar-diameter}
\end{figure}
\end{example}
\vspace*{-8mm}
Our purpose is to transform the above centralized algorithm into a
distributed one while satisfying the complexity bound
$MSG/LINK=O(1)$. The direct transformation will imply that we should
\vspace*{-2mm}
\begin{itemize}
\item first triangulate distributively the planar network,

\item then construct
distributively a BFS tree for the triangulated network,

\item finally construct
and store distributively the ordered tree decomposition by using the
BFS tree,
\end{itemize}
\vspace*{-2mm}
while ensuring the complexity bound $MSG/LINK=O(1)$.

Nevertheless, the direct transformation seems infeasible: Even if we
can triangulate the biconnected planar network within the complexity
bound, it is difficult to construct a BFS tree for the triangulated
network with only $O(1)$ messages sent over each link, because the
triangulated network includes \emph{virtual} links between nodes,
and two nodes connected by a virtual link may be far away from each
other in the real network.

A key observation to tackle this difficulty is that in the above
centralized algorithm, a tree decomposition of $G$ can be obtained
by using \emph{any spanning tree} of $G^\prime$, not necessarily a
BFS tree of it. Thus we can construct a BFS tree of $G$, instead of
$G^\prime$, which can be done with only $O(1)$ messages sent over
each link \cite{BDLP08}, and use it to construct the tree
decomposition.

The distributed algorithm to construct an ordered tree decomposition
for biconnected planar networks with bounded diameter works as
follows:
\vspace*{-2mm}
\begin{itemize}
\item A BFS-tree of the planar network with the requesting node as the root is distributively constructed and stored;

\item A post-order traversal of the BFS-tree can be done, those facial cycles
are visited one by one, all the faces in the combinatorial
embedding, including the outer face, are triangulated, and the bags
corresponding to the triangles are stored among the nodes of the
network;

\item Finally, some bag stored in the requesting node can be selected as the root bag, and
the bags are connected together depending on whether the
corresponding triangles share a non-BFS-tree link or not.
\end{itemize}
\vspace*{-2mm}
We now describe more specifically the post-order
traversal of the BFS tree, and how to connect the distributively
stored bags into a tree decomposition.

Each link $\{v,w\}$ is seen as two arcs $(v,w)$ and $(w,v)$. Let $l$
be a port of node $v$, then $neighbor(l)$ denotes the neighbor of
$v$ corresponding to the port $l$.

\smallskip
\noindent \textbf{Post-order traversal of the BFS tree}.

A post-order traversal of the BFS tree is done to visit the nodes
one by one.

When a node $v$ is traversed,

If there are arcs $(v,w)$ not visited, let
%
\[l_0:=\min \{ l | \exists w \ s.t. \ (v,w) \mbox{ is not visited } \mbox{ and } neighbor(l)=w \}\]
%
Let $w_0=neighbor(l_0)$, and start traversing the facial cycle which
the arc $(v,w_0)$ belongs to ($(v,w_0)$ is called the starting arc
of the facial cycle, and is seen as the identifier of the facial
cycle), by using the cyclic order in each node.

The facial cycle with $(v,w_0)$ as the starting arc will be
triangulated by virtually connecting $v$ to all the non-neighbor
nodes of $v$ in the facial cycle. When an arc $(v^\prime,w^\prime)$
such that $w^\prime \ne v$ and $w^\prime \ne w_0$ in the facial
cycle is visited, the bag (in the ordered tree decomposition)
corresponding to the triangle $[v,v^\prime,w^\prime]$ will be stored
in $w^\prime$. In addition, $w_0$ stores the bag $[v,w_0,w_1]$,
where $w_1$ is the node visited immediately after $w_0$ in the
facial cycle; and $v$ stores the bag $[v,v^\ast,w^\ast]$, where
$v^\ast$,$w^\ast$ are the last two nodes visited in the facial
cycle.

When the traversal of a facial cycle is finished, i.e. $v$ is
reached again during the traversal of the facial cycle, then repeat
the above procedure until all the arcs $(v,w)$ are visited.

When all the arcs $(v,w)$ have been visited, backtrack to the parent
of $v$ in the BFS tree.

\smallskip

\noindent \textbf{Connect the bags into a tree decomposition}.

If a bag $[v^\prime,v^{\prime\prime},v^{\prime\prime\prime}]$ is
stored on a node $v$ during the above traversal of a facial cycle,
then $[v^\prime,v^{\prime\prime},v^{\prime\prime\prime}]$ is said to
be the bag stored on $v$ corresponding to the facial cycle.

First, select some bag
$[v^\prime,v^{\prime\prime},v^{\prime\prime\prime}]$ stored in the
requesting node as the root of the ordered tree decomposition.

Then start visiting the bags stored on the nodes in the facial cycle
which has $(v^\prime,v^{\prime\prime})$ as the starting arc. Now we
describe how to visit and connect the bags into a tree
decomposition.

Let $(w_0,\cdots, w_m)$ be the facial cycle currently visited
($w_0,w_m$ are resp. the first and last visited node during the
virtual triangulation process above), and
$[v^\prime_0,v^{\prime\prime}_0,v^{\prime\prime\prime}_0]$,
$\cdots$, $[v^\prime_m,v^{\prime\prime}_m,v^{\prime\prime\prime}_m]$
be resp. the bags stored on $w_0,\cdots,w_m$ corresponding to the
facial cycle.

Suppose $[v^\prime_l,v^{\prime\prime}_l,v^{\prime\prime\prime}_l]$
($0 \le l \le m$) is the first visited bag among them during this
bag-connecting process.

Let $w_{m+1}=w_0$ by convention.

If $l=0$, then the nodes in the facial cycle will be visited
according to the order $w_0 w_m \cdots w_1$. So
$[v^\prime_{i+1},v^{\prime\prime}_{i+1},v^{\prime\prime\prime}_{i+1}]$
is taken as the father of
$[v^\prime_{i},v^{\prime\prime}_{i},v^{\prime\prime\prime}_{i}]$ for
all $1 \le i \le m$ in the tree decomposition.

Note that above we let
$[v^\prime_m,v^{\prime\prime}_m,v^{\prime\prime\prime}_m]$ and
$[v^\prime_0,v^{\prime\prime}_0,v^{\prime\prime\prime}_0]$
\emph{stay together} in the tree decomposition because the content
of the two bags are in fact the same.

If $l=1$, then the nodes in the facial cycle are visited according
to the order $w_1 \cdots w_m w_0$. So
$[v^\prime_i,v^{\prime\prime}_i,v^{\prime\prime\prime}_i]$ is taken
as the father of
$[v^\prime_{i+1},v^{\prime\prime}_{i+1},v^{\prime\prime\prime}_{i+1}]$
for all $1 \le i \le m$ in the tree decomposition.

Otherwise ($l > 1$), then the nodes in the facial cycle are visited
concurrently along the two lines according to the order $w_l w_{l+1}
\cdots w_m w_0$ and $w_l w_{l-1} \cdots w_1$ respectively. So
$[v^\prime_i,v^{\prime\prime}_i,v^{\prime\prime\prime}_i]$ is taken
as the father of
$[v^\prime_{i+1},v^{\prime\prime}_{i+1},v^{\prime\prime\prime}_{i+1}]$
for all $l \le i \le m$,
$[v^\prime_{j+1},v^{\prime\prime}_{j+1},v^{\prime\prime\prime}_{j+1}]$
is taken as the father of
$[v^\prime_j,v^{\prime\prime}_j,v^{\prime\prime\prime}_j]$ for all
$l > j \ge 1$ in the tree decomposition.

Moreover, if during the above process, a node $w_{i+1}$ ($1 \le i
\le m$) is visited through an arc $(w_i,w_{i+1})$ in the facial
cycle, and $\{w_i,w_{i+1}\}$ is a non-BFS-tree link, then start
visiting the new facial cycle which $(w_{i+1},w_i)$ belongs to; on
the other hand, if $w_{i}$ ($i > 1 $) is visited, and the arc
$(w_i,w_{i-1})$ will be visited next, moreover $\{w_i,w_{i-1}\}$ is
a non-BFS-tree link, then start visiting the new facial cycle which
$(w_{i},w_{i-1})$ belongs to.

The detailed distributed algorithm is given in the appendix.

It is not hard to see that only $O(1)$ messages are sent over each
link during the computation of the above distributed algorithm. Then
we get the following lemma.

\vspace*{-2mm}
\begin{lemma}
An ordered tree decomposition of biconnected planar networks with
bounded diameter can be distributively constructed within the
complexity bounds $TIME=O(n)$ and $MSG/LINK=O(1)$.
\end{lemma}
\vspace*{-6mm}
\begin{example}
[Distributed ordered tree decomposition of biconnected planar
networks with bounded diameter] A biconnected planar network is
given in Fig.\ref{fig-triangulation-biconn-planar-diameter}(a) with
thick lines denoting the edges of the distributively constructed
BFS-tree rooted on the requesting node $1$. The cyclic ordering of
each node is the same as the order of the identifiers of the ports.
During the post-order traversal of the BFS tree, node $3$ is first
traversed, then $2$, $8$, $7$, $6$, $9$, $4$, $5$ and finally $1$.
The facial cycles corresponding to face $A$ is visited first, then
$B$, $E$, $C$, $D$. The triangulation after the post-order traversal
of the BFS-tree is given in
Fig.\ref{fig-triangulation-biconn-planar-diameter}(b). The
constructed tree decomposition is illustrated in
Fig.\ref{fig-triangulation-biconn-planar-diameter}(c). Note that in
Fig.\ref{fig-triangulation-biconn-planar-diameter}(c), we only give
the distributively stored bags corresponding to face $A$ and $B$,
and omit the others in order to avoid overfilling the figure.
Suppose the bag $[3,7,1]$ is selected as the root of the tree
decomposition, then the child-parent relationship between these
distributively stored bags are illustrated in
Fig.\ref{fig-triangulation-biconn-planar-diameter}(c).

\vspace*{-6mm}
\begin{figure}[ht]
\centering
  \includegraphics[width=\textwidth]{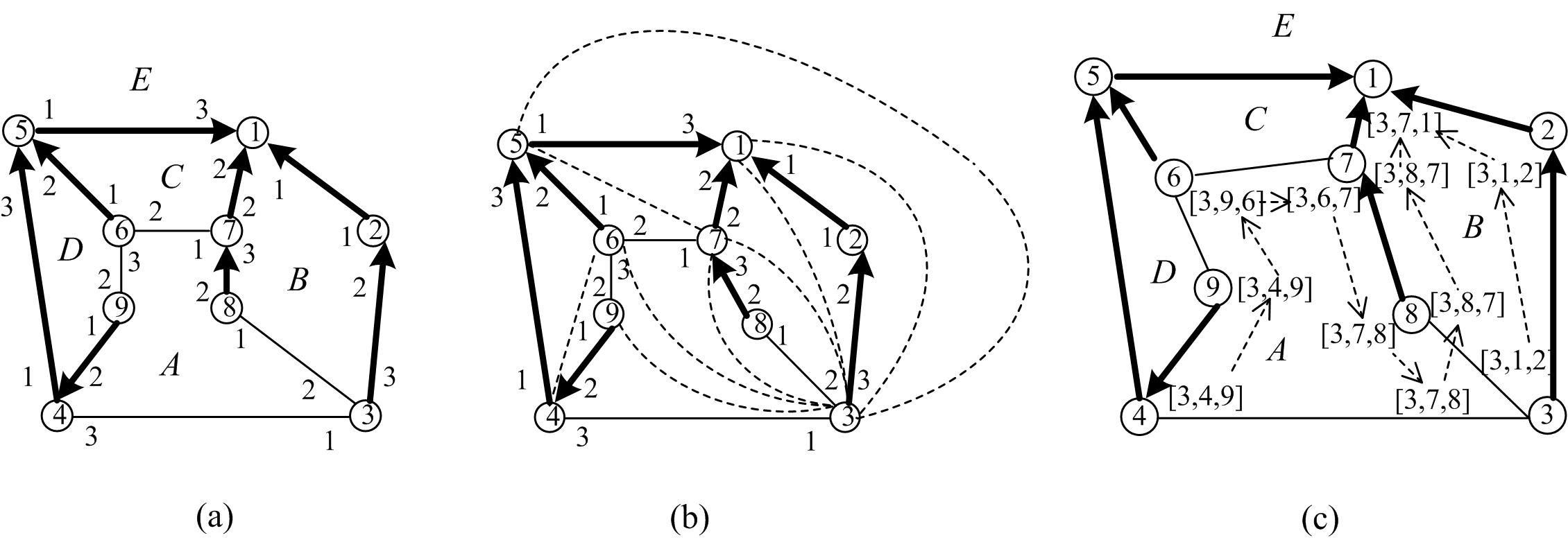}\\
  \caption{Distributed ordered tree decomposition of biconnected planar networks with bounded diameter}\label{fig-triangulation-biconn-planar-diameter}
\end{figure}
\vspace*{-8mm}
\end{example}

\vspace*{-5mm}
\subsection{General planar networks with bounded diameter}

Now we consider the general case when the planar networks with
bounded diameter are not necessarily biconnected. We first state a
proposition on the relationship between the spanning tree of a given
graph and the spanning trees of its biconnected components.
\vspace*{-2mm}
\begin{proposition}\label{prop-BFS-tree-biconnected}
Let $G=(V,E)$ be a connected graph, $T$ a spanning tree of $G$, and
$B=(V_B,E_B)$ be a nontrivial biconnected component of $G$, then
$T[V_B]$, the subgraph of $T$ induced by $V_B$, is a spanning tree
of $B$.
\end{proposition}
\vspace*{-2mm}
To construct distributively the ordered tree decomposition for
(general) planar networks with bounded diameter, we first compute a
BFS tree with only $O(1)$ messages sent over each link, and compute
distributively the biconnected components of the network by using
the algorithm given in \cite{Turau06} within the complexity bounds
$TIME=O(n)$ and $MSG/LINK=O(1)$. Then let $T$ be the computed
BFS-tree of the network, we compute separately the ordered tree
decomposition for each biconnected component $B=(V_B,E_B)$ by using
$T[V_B]$, the subtree of $T$ induced by $V_B$. Finally these ordered
tree decompositions are connected together into a complete ordered
tree decomposition of the whole network.
\vspace*{-2mm}
\begin{lemma}
The distributed construction of an ordered tree decomposition for
(general) planar networks with bounded diameter can be done within
the complexity bounds $TIME=O(n)$ and $MSG/LINK=O(1)$.
\end{lemma}
\vspace*{-6mm}

\section{Towards (general) networks with bounded tree
width}\label{sec-MSO-BTL}
\vspace*{-2mm}
In the last section, we have shown that MSO can be distributively
model-checked over planar networks with bounded diameter with only
$O(1)$ messages sent over each link. Courcelle's classical result
states that MSO can be model-checked on graphs with bounded tree
width in linear time. Then a natural question to ask is whether we
can extend Theorem~\ref{thm-MSO-sentence-planar-diameter} to the
(general) networks with bounded tree width.


In the centralized linear-time construction of the tree
decomposition, distances between nodes are usually ignored, and two
vertices contained in the same bag of the tree decomposition may be
far away from each other in the original graph. Thus it seems in
general quite difficult, if not impossible, to transform the
linear-time centralized tree decomposition algorithm into the
distributed one with only a constant number of messages sent over
each link. As a matter of fact, distances between vertices in the
centralized tree decomposition have been considered in \cite{DG04},
where the concept of tree length of a tree decomposition, which is
the maximal distance (in the original graph) between vertices in the
same bag of a tree decomposition, was defined and investigated.

In this section, based on the work that has been done in
\cite{DG04}, we consider the distributed model-checking of MSO over
networks with bounded degree and bounded tree-length. These classes
of networks are of independent interest and they have been applied
to construct compact routing schemes for forwarding messages
\cite{Dour04}. However, even for these networks, we can only achieve
the complexity bound $MSG/LINK=O(1)$ over two more restricted
models, namely \emph{synchronous} distributed systems and
\emph{asynchronous} distributed systems with \emph{a BFS tree
pre-computed and distributively stored} on each node of the network
(each node stores locally its parent in the BFS tree). The reason
for this restriction on the computational models is in that a BFS
tree is essential for the distributed tree decomposition, and
currently we do not know how to distributively construct a BFS tree
in asynchronous systems with only $O(1)$ messages sent over each
link, and the best complexity bound achieved is $MSG/LINK=O(\Delta)$
\cite{BDLP08}.

Let $G=(V,E)$ and $X \subseteq V$, then the diameter of $X$ in $G$,
denoted $diam_G(X)$, is defined by $\max \{dist_G(v,w) | v,w \in
X\}$. Let $\mathcal{T}=(I,F,r,B)$ be a tree decomposition of $G$,
then the length of $\mathcal{T}$, $length(\mathcal{T})$, is defined
by $\max\{diam_G(B(i))|i \in I\}$. The \emph{tree-length} of $G$,
denoted $tl(G)$, is the minimum length over all tree decompositions
of $G$.

Let $\mathcal{C}$ be a class of graphs, we say that $\mathcal{C}$
has bounded tree-length if there is a constant $k$ such that for
each $G \in \mathcal{C}$, $tl(G) \le k$.
%
%
The following proposition is easy to verify.
\vspace*{-1.5mm}
\begin{proposition}
Let $\mathcal{C}$ be a class of networks of bounded degree and
bounded tree-length, then $\mathcal{C}$ has bounded tree-width.
\end{proposition}
\vspace*{-1.5mm}
In the rest of this subsection, we assume that each node of the
network stores locally a bound $d$ on the degree, and a bound $k$ on
the tree-length.

%
%
\vspace*{-2mm}
\begin{theorem}\label{thm:MSO-BTL-frugal}
MSO can be distributively model-checked over networks with bounded
degree and bounded tree-length within the complexity bounds
$TIME=O(\Delta)$ and $MSG/LINK=O(1)$, in the following two
computational models,
\vspace*{-1.5mm}
\begin{itemize}
\item synchronous distributed systems,
\item or asynchronous distributed systems with a BFS tree
pre-computed and distributively stored on each node of the network.
\end{itemize}
\vspace*{-1.5mm}
\end{theorem}
\vspace*{-3mm}
%

The main idea of the proof is to distributively construct an ordered
tree decomposition by transforming the BFS-layering tree
decomposition algorithm in \cite{DG04}. In the following, we give a
more specific description of the construction.

We first recall the centralized construction of the BFS-layering
tree decomposition.

Let $s$ be a distinguished vertex in graph $G=(V,E)$. Let $L^i=\{v
\in V | dist_G(s,v) = i\}$. A \emph{layering partition} of $G$ is a
partition of each set $L^i$ into $L^i_1,\cdots, L^i_{p_i}$ such that
$u,v \in L^i_j$ if and only if there exists a path from $u$ to $v$
such that all the intermediate vertices $w$ on the path satisfy that
$dist_G(s,w) \ge i$.

Let $H=(V_H,E_H)$ be the graph defined as follows:

\vspace*{-1.5mm}
\begin{itemize}
\item $V_H$: the sets $L^i_j$.
\vspace*{-.5mm}
\item $E_H$: $\{L^i_j,L^{i^\prime}_{j^\prime}\} \in E_H$ if and only
if there is $u \in L^i_j$ and $v \in L^{i^\prime}_{j^\prime}$ such
that $(u,v) \in E$.
\end{itemize}
\vspace*{-5mm}
\begin{theorem}
\cite{DG04} The graph $H$ is a tree.
\end{theorem}
\vspace*{-2mm}

$H$ is called the \emph{BFS-layering tree} of $G$.

$H$ can be seen as a rooted labeled tree $(I,F,r,B)$ with
\vspace*{-2mm}
\begin{itemize}
\item $I=\{(i,j)| L^i_j \in V_H\}$, $F=\{\{(i,j),(i^\prime,j^\prime)\} |
\{L^i_j,L^{i^\prime}_{j^\prime}\} \in E_H\}$,

\item The root $r$ is $(0,1)$,

\item $B((i,j))=L^i_j$.
\end{itemize}
\vspace*{-2mm}

If we replace the label of $(i,j)$ by $B(i,j) \cup
B(i^\prime,j^\prime)$ in $H$, where $(i^\prime,j^\prime)$ is the
parent of $(i,j)$ in $H$, then we get a new rooted labeled tree
$\mathcal{S}=(I,F,r,L)$.

\vspace*{-2mm}
\begin{theorem}\label{thm:BFS-layering-tree-decomp}
\cite{DG04} $\mathcal{S}$ is a tree decomposition of $G$ such that
$length(\mathcal{S}) \le 3 \cdot tl(G)+1$.
\end{theorem}
\vspace*{-2mm}

$\mathcal{S}$ is called the \emph{BFS-layering tree decomposition}
of $G$.

\noindent Now we transform the above centralized construction into a
distributed algorithm in asynchronous distributed systems with a
pre-computed BFS tree. The asynchronous distributed algorithm
consists of two stages:

\vspace*{-1.5mm}
\begin{description}
\item[Stage 1]: Construct the BFS-layering tree $H$
bottom-up. Because $G$ has bounded tree-length $t$, $\mathcal{S}$
has length no more than $3t+1$ by
Theorem~\ref{thm:BFS-layering-tree-decomp}. Thus if two nodes of
layer $L^i$ are in the same layering partition, then the distance
between them is no more than $3t+1$. Consequently, when the layering
partition of $L^{i+1}$ has been computed, each node $v$ in $L^i$ can
know which nodes are in the same layering partition of $L^i$ by
doing some local computation in its $(3t+1)$-neighborhood.
%
\item [Stage 2]: Construct the BFS-layering tree decomposition $\mathcal{S}$ from $H$, and an ordered tree decomposition $\mathcal{T}$ from $\mathcal{S}$.
\end{description}
\vspace*{-1.5mm}

The distributed algorithm to construct the ordered tree
decomposition in synchronous systems is similar to the above
two-stage algorithm, except that a stage for the BFS-tree
construction should be added before the above two stages.

\smallskip

The tree-length of (general) networks with bounded tree-width may be
unbounded. The tree-width of a cycle of length $n$ for instance is
$2$, while its tree-length is $\lceil n/3 \rceil$ \cite{DG04}. For
networks of unbounded tree-length, vertices in the same bag of a
tree decomposition of the network may be arbitrarily far away from
each other. The above technique we use to construct and store the
tree decomposition within the complexity bound $MSG/LINK=O(1)$
doesn't carry over to unbounded tree-length networks, and it seems
difficult to extend it.

\vspace*{-3mm}

\section{Conclusion}

\vspace*{-2mm} We have seen that MSO sentences can be distributively
model-checked over classes of planar networks with bounded diameter,
as well as classes of networks with bounded degree and bounded
tree-length, with only constant number of messages sent over each
link.


So far as the class of graphs on which the results hold is
concerned, we doubt that our techniques for MSO can be extended to
bounded tree-width graphs, but we were able to prove the result for
$k$-outerplanar graphs.

Similar to the centralized computation, the expression complexity
for MSO distributed model checking is very high, which hinders the
practical value of results obtained in this paper. However, we can
relieve the difficulty to some extent by encoding symbolically the
tree automata obtained from MSO sentences, in Binary Decision
Diagrams (BDD), just like in the MSO model-checker MONA
\cite{HJJ+96}.

\vspace*{-3mm}

\small
\bibliographystyle{alpha}
\bibliography{bibliolocal}

\newpage

\normalsize

\addtolength{\hoffset}{-3cm} \addtolength{\textwidth}{5cm}

\addtolength{\voffset}{-1cm} \addtolength{\textheight}{3cm}

\onecolumn

\begin{appendix}

\section{Step 2 and Step 3 of the linear-time MSO model checking over graphs with bounded tree width}

For Step 2, a rooted $\Sigma_{k}$-labeled tree
$\mathcal{T}^\prime=(T,F,r,\lambda)$, where $\Sigma_k=2^{[k+1]^2}
\times 2^{[k+1]^2} \times 2^{[k+1]^2}$ ($[k+1]=\{1,2,\cdots,k+1\}$),
can be obtained from $\mathcal{T}$ as follows: The new labeling
$\lambda$ over $(T,F)$ is defined by
$\lambda(t)=(\lambda_1(t),\lambda_2(t),\lambda_3(t))$, where
\vspace*{-2mm}
\begin{itemize}
\item $\lambda_1(t):=\{(j_1,j_2)\in[k+1]^2 | (b^t_{j_1},b^t_{j_2}) \in
E\}$.

\item $\lambda_2(t):=\{(j_1,j_2)\in[k+1]^2 | b^t_{j_1}=b^t_{j_2}\}$.

\item $\lambda_3(t):=
\left\{\begin{array}{cc} \{(j_1,j_2)\in[k+1]^2|
b^t_{j_1}=b^{t^\prime}_{j_2}\} &
    \mbox{for the parent } t^\prime \mbox{ of } t, \mbox{ if } t \ne r\\
\emptyset & \mbox{if } t = r
\end{array} \right.$
\end{itemize}
\vspace*{-2mm}
For Step 3, we recall how to translate the MSO sentence $\varphi$
over the vocabulary $\{E\}$ into an MSO sentence $\varphi^\ast$ over
the vocabulary $\{E_1,E_2\} \cup \{P_c| c \in \Sigma_k\}$ such that
$G \models \varphi$ iff $\mathcal{T}^\prime \models \varphi^\ast$.
The translation relies on the observation that elements  and subsets
of $V$ can be represented by  $(k+1)$-tuples of subsets of $T$. For
each element $v \in V$ and $i \in [k+1]$, let
\[
U_i(v):=\left\{\begin{array}{cl}
            \{t(v)\} & \mbox{, if }b^{t(v)}_i=v\mbox{, and }b^{t(v)}_j \ne v \mbox{ for all }j: 1 \le j < i \\
            \emptyset & \mbox{, otherwise}
          \end{array}\right.
\]
where $t(v)$ is the minimal $t \in T$ (with respect to the partial
order $\le^{\mathcal{T}}$) such that $v \in
\{b^{t}_1,\cdots,b^{t}_{k+1}\}$. Let
$\overline{U}(v)=(U_1(v),\cdots,U_{k+1}(v))$.

For each $S \subseteq V$ and $i \in [k+1]$, let $U_i(S):=\cup_{v \in
S} U_i(v)$, and let $\overline{U}(S)=(U_1(S),\cdots,U_{k+1}(S))$. It
is not hard to see that for subsets $U_1,\cdots,U_{k+1} \subseteq
T$, there exists $v \in V$ such that $\overline{U}=\overline{U}(v)$
iff
\vspace*{-2mm}
\begin{itemize}
\item (1) $\bigcup^{k+1}_{i=1} U_i$ is a singleton;

\item (2) For all $t \in T$, $i<j<k+1$, if $t \in U_j$, then $(i,j) \not \in \lambda_2(t)$;

\item (3) For all $t \in T$, $i,j<k+1$, if $t \in U_i$, then $(i,j) \not \in
\lambda_3(t)$.
\end{itemize}
\vspace*{-2mm}
\noindent Moreover, there is a subset $S \subseteq V$ such that
$\overline{U}=\overline{U}(S)$ iff conditions (2) and (3) are
satisfied. Using the above characterizations of $\overline{U}(v)$
and $\overline{U}(S)$, it is easy to construct MSO formulas
$Elem(X_1,\cdots,X_{k+1})$ and $Set(X_1,\cdots,X_{k+1})$ over
$\{E_1,E_2\} \cup \{P_c| c \in \Sigma_k\}$ such that

$\mathcal{T}^\prime \models Elem(\overline{U}) \mbox{ iff there is a
} v \in V \mbox{ such that } \overline{U}=\overline{U}(v).$

$\mathcal{T}^\prime \models Set(\overline{U}) \mbox{ iff there is a
} S \subseteq V \mbox{ such that } \overline{U}=\overline{U}(S).$

\vspace*{-2mm}
\begin{lemma}\label{lem-MSO-translation}
\cite{FFG02} Every MSO formula
$\varphi(X_1,\cdots,X_r,y_1,\cdots,y_s)$ over vocabulary $E$ can be
effectively translated into a formula
$\varphi^\ast(\overline{X}_1,\cdots,\overline{X}_r,\overline{Y}_1,\cdots,\overline{Y}_s)$
over the vocabulary $\{E_1,E_2\} \cup \{P_c | c \in \Sigma_k\}$ such
that
\vspace*{-2mm}
\begin{description}
\item (1) For all $S_1,\cdots,S_r \subseteq V$, and $v_1,\cdots,v_s \in V$,
\[G \models \varphi(S_1,\cdots,S_r,v_1,\cdots,v_s) \mbox{ iff }
\mathcal{T}^\prime \models
\varphi^\ast(\overline{U}(S_1),\cdots,\overline{U}(S_r),\overline{U}(v_1),\cdots,\overline{U}(v_s)).\]

\item (2) For all $\overline{U}_1,\cdots,\overline{U}_r,\overline{W}_1,\cdots,\overline{W}_s \subseteq T$ such
that $\mathcal{T}^\prime \models
\varphi^\ast(\overline{U}_1,\cdots,\overline{U}_r,\overline{W}_1,$
$\cdots,\overline{W}_s)$, there exist $S_1,\cdots,S_r \subseteq V$,
$v_1,\cdots,v_s \in V$ such that $\overline{U}_i=\overline{U}(S_i)$
for all $1\le i \le r$ and $\overline{W}_j=\overline{U}(v_j)$ for
all $1 \le j \le s$.
\end{description}
\vspace*{-3mm}
\end{lemma}


\section{Distributed model-checking of MSO over tree networks}\label{appendix-MSO-tree}

Suppose each node stores the states of ports, ``parent'' or
``child'', with respect to the rooted tree (with the requesting node
as the root).


Through local replacement of each node $v$ by the set of (virtual)
nodes $\{[v,i]| 1\le i \le deg(v)\}$, the network is first
transformed into a (virtual) binary tree, then an ordered tree
decomposition $\mathcal{T}$ of width $1$ and rank $\le 2$ is
obtained;

The tree decomposition $\mathcal{T}$ is transformed into a
$\Sigma_2$-labeled binary tree $\mathcal{T}^\prime$;

The requesting node constructs a tree automata $\mathcal{A}$ from
$\varphi$, and broadcasts $\mathcal{A}$ to all the nodes in the
network;

Finally $\mathcal{A}$ is ran distributively over
$\mathcal{T}^\prime$ in a bottom-up way to decide whether
$\mathcal{T}^\prime$ is accepted by $\mathcal{A}$.

In the following we describe the distributed algorithm in detail by
giving the pseudo-code for the message processing at each node $v$.

\medskip
\noindent
  \begin{tabular*}{\textwidth}{l@{\extracolsep{\fill}}}
    \hline
    Initialization \\
    \hline
    The requesting node $v_0$ sets $Virtual(v_0):=\langle
    [v_0,1],[v_0,2],\dots,[v_0,deg(v_0)]\rangle$.\\
    The requesting node sends message START over all its ports.\\
    \hline
    \\
    \hline
    Message START over port $l$\\
    \hline
    $Virtual(v):=\langle [v,1],[v,2],\dots,[v,deg(v)]\rangle$.\\
    \textbf{if} $deg(v) > 1$ \textbf{then}\\
    \hspace*{1em} $v$ sends message START over all ports $l^\prime$ such that $state(l^\prime)=$``child''.\\
    \textbf{else}\\
    \hspace*{1em} $v$ sends message ACK over port $l$.\\
    \textbf{end if}\\
    \hline
    \\
    \hline
    Message ACK over port $l$\\
    \hline
    $bAck(l):=true$.\\
    \textbf{if} $bAck(l^\prime)=true$ for each port $l^\prime$ such that $state(l^\prime)=$``child'' \textbf{then}\\
    \hspace*{1em} \textbf{if} $v$ is the requesting node \textbf{then}\\
    \hspace*{1em}\hspace*{1em} $v$ sends message TREEDECOMP over all its ports.\\
    \hspace*{1em} \textbf{else}\\
    \hspace*{1em}\hspace*{1em} $v$ sends message ACK over the port $l^\prime$ such that $state(l^\prime)=$``parent''.\\
    \hspace*{1em} \textbf{end if}\\
    \textbf{end if}\\
    \hline
    \\
    \hline
    Message TREEDECOMP over port $l$\\
    \hline
    $Bag([v,l]):=(v,parent(v))$. 
    \textbf{for each} $l^\prime: 1 \le l^\prime \le deg(v), l^\prime \ne
    l$ \textbf{do}\\
    \hspace*{1em} $Bag([v,l^\prime]):=(v,v)$.\\
    \textbf{end for}\\
    \textbf{if} $deg(v)=1$ \textbf{then}\\
    \hspace*{1em} $v$ sends message DECOMPOVER over the port $l^\prime$ such that $state(l^\prime)=$``parent''.\\
    \textbf{else}\\
    \hspace*{1em} $v$ sends message TREEDECOMP over all ports $l^\prime$ such that $state(l^\prime)=$``child''.\\
    \textbf{end if}\\
    \hline
    \\
    \hline
    Message DECOMPOVER over port $l$\\
    \hline
    $bDecompOver(l):=true$.\\
    \textbf{if} $bDecompOver(l^\prime)=true$ for each port $l^\prime$ such that $state(l^\prime)=$``child'' \textbf{then} \\
    \hspace*{1em} \textbf{if} $v$ is the requesting node \textbf{then}\\
    \hspace*{1em}\hspace*{1em} \textbf{for each} $l^\prime$ \textbf{do}\\
    \hspace*{1em}\hspace*{1em}\hspace*{1em} $\lambda_1([v,l^\prime]):=\emptyset$, $\lambda_2([v,l^\prime]):=\{(1,2),(2,1)\}$, $\lambda_3([v,l^\prime]):=\emptyset$, $\lambda([v,l^\prime]):=(\lambda_1([v,l^\prime]),\lambda_2([v,l^\prime]),\lambda_3([v,l^\prime]))$.\\
    \hspace*{1em}\hspace*{1em}\hspace*{1em} $v$ sends message TREELABELING over port $l^\prime$.\\
    \hspace*{1em}\hspace*{1em} \textbf{end for}\\
    \hspace*{1em} \textbf{else}\\
    \hspace*{1em}\hspace*{1em} $v$ sends message DECOMPOVER over the port $l^\prime$ such that $state(l^\prime)=$``parent''.\\
    \hspace*{1em} \textbf{end if}\\
    \textbf{end if}\\
    \hline
  \end{tabular*}

\noindent
\begin{tabular*}{\textwidth}{l@{\extracolsep{\fill}}}
    \hline
    Message TREELABELING over port $l$\\
    \hline
    $\lambda_1([v,l]):=\{(1,2),(2,1)\}$, $\lambda_2([v,l]):=\emptyset$, $\lambda_3([v,l]):=\{(2,1),(2,2)\}$, $\lambda([v,l]):=(\lambda_1([v,l]),\lambda_2([v,l]),\lambda_3([v,l]))$.\\
    \textbf{for each} $l^\prime: 1 \le l^\prime \le deg(v),l^\prime \ne l$ \textbf{do}\\
    \hspace*{1em} $\lambda_1([v,l^\prime]):=\emptyset$, $\lambda_2([v,l^\prime]):=\{(1,2),(2,1)\}$.\\
    \textbf{end for}\\
    \textbf{if} $deg(v)>1$ \textbf{then}\\
    \hspace*{1em} \textbf{if} $l=1$ \textbf{then}\\
    \hspace*{1em}\hspace*{1em} $\lambda_3([v,2]):=\{(1,1),(2,1)\}$, $\lambda([v,2]):=(\lambda_1([v,2]),\lambda_2([v,2]),\lambda_3([v,2]))$.\\
    \hspace*{1em}\hspace*{1em} \textbf{for each} $l^\prime: 3 \le l^\prime \le deg(v)$ \textbf{do}\\
    \hspace*{1em}\hspace*{1em}\hspace*{1em} $\lambda_3([v,l^\prime]):=\{(1,1),(1,2),(2,1),(2,2)\}$, $\lambda([v,l^\prime]):=(\lambda_1([v,l^\prime]),\lambda_2([v,l^\prime]),\lambda_3([v,l^\prime]))$.\\
    \hspace*{1em}\hspace*{1em} \textbf{end for}\\
    \hspace*{1em} \textbf{else if} $l=deg(v)$ \textbf{then}\\
    \hspace*{1em}\hspace*{1em} $\lambda_3([v,deg(v)-1]):=\{(1,1),(2,1)\}$.\\
    \hspace*{1em}\hspace*{1em} $\lambda([v,deg(v)-1]):=(\lambda_1([v,deg(v)-1]),\lambda_2([v,deg(v)-1]),\lambda_3([v,deg(v)-1]))$.\\
    \hspace*{1em}\hspace*{1em} \textbf{for each} $l^\prime: 1 \le l^\prime \le deg(v)-2$ \textbf{do}\\
    \hspace*{1em}\hspace*{1em}\hspace*{1em} $\lambda_3([v,l^\prime]):=\{(1,1),(1,2),(2,1),(2,2)\}$, $\lambda([v,l^\prime]):=(\lambda_1([v,l^\prime]),\lambda_2([v,l^\prime]),\lambda_3([v,l^\prime]))$.\\
    \hspace*{1em}\hspace*{1em} \textbf{end for}\\
    \hspace*{1em} \textbf{else}\\
    \hspace*{1em}\hspace*{1em} $\lambda_3([v,l-1]):=\lambda_3([v,l+1]):=\{(1,1),(2,1)\}$.\\
    \hspace*{1em}\hspace*{1em} $\lambda([v,l-1]):=(\lambda_1([v,l-1]),\lambda_2([v,l-1]),\lambda_3([v,l-1]))$.\\
    \hspace*{1em}\hspace*{1em} $\lambda([v,l+1]):=(\lambda_1([v,l+1]),\lambda_2([v,l+1]),\lambda_3([v,l+1]))$.\\
    \hspace*{1em}\hspace*{1em} \textbf{for each} $l^\prime: 1 \le l^\prime \le deg(v), l^\prime \ne l-1,l, l+1$ \textbf{do}\\
    \hspace*{1em}\hspace*{1em}\hspace*{1em} $\lambda_3([v,l^\prime]):=\{(1,1),(1,2),(2,1),(2,2)\}$, $\lambda([v,l^\prime]):=(\lambda_1([v,l^\prime]),\lambda_2([v,l^\prime]),\lambda_3([v,l^\prime]))$.\\
    \hspace*{1em}\hspace*{1em} \textbf{end for}\\
    \hspace*{1em} \textbf{end if}\\
    \hspace*{1em} $v$ sends message TREELABELING over each port $l^\prime$ such that $state(l^\prime)=$``child''.\\
    \textbf{else}\\
    \hspace*{1em} $v$ sends message LABELINGOVER over the port $l^\prime$ such that $state(l^\prime)=$``parent''.\\
    \textbf{end if}\\
    \hline
    \\
    \hline
    Message LABELINGOVER over port $l$\\
    \hline
    $bLabelingOver(l):=true$.\\
    \textbf{if} $bLabelingOver(l^\prime)=true$ for each port $l^\prime$ such that $state(l^\prime)=$``child'' \textbf{then}\\
    \hspace*{1em} \textbf{if} $v$ is the requesting node \textbf{then}\\
    \hspace*{1em}\hspace*{1em} $v$ constructs tree automaton $\mathcal{A}=(Q,\Sigma_2,\delta,f_0,F)$ from $\varphi$.\\
    \hspace*{1em}\hspace*{1em} $v$ sends message AUTOMATON($\mathcal{A}$) over all its ports.\\
    \hspace*{1em} \textbf{else}\\
    \hspace*{1em}\hspace*{1em} $v$ sends message LABELINGOVER over the port $l^\prime$ such that $state(l^\prime)=$``parent''.\\
    \hspace*{1em} \textbf{end if}\\
    \textbf{end if}\\
    \hline
  \end{tabular*}

\noindent
    \begin{tabular*}{\textwidth}{l@{\extracolsep{\fill}}}
    \hline
    Message AUTOMATON($\mathcal{A}$) over port $l$\\
    \hline
    Let $\mathcal{A}=(Q,\Sigma_2,\delta,f_0,F)$.\\
    $automaton(v):=(Q,\Sigma_2,\delta,f_0,F)$.\\
    \textbf{if} $deg(v)=1$ \textbf{then}\\
    \hspace*{1em} $v$ sends message STATE($f_0(\lambda([v,1]))$) over the port $l^\prime$ such that $state(l^\prime)=$``parent''.\\
    \textbf{else}\\
    \hspace*{1em} $v$ sends message AUTOMATON($\mathcal{A}$) over all ports $l^\prime$ such that $state(l^\prime)=$``child''.\\
    \textbf{end if}\\
    \hline
    \\
    \hline
    Message STATE($q$) over port $l$\\
    \hline
    $bState(l):=true$, $childState(l):=q$.\\
    \textbf{if} $bState(l^\prime)=true$ for each port $l^\prime$ such that $state(l^\prime)=$``child'' \textbf{then} \\
    \hspace*{1em} \textbf{if} $v$ is the requesting node \textbf{then}\\
    \hspace*{1em}\hspace*{1em} $state([v,deg(v)]):=\delta(childState(deg(v)),\lambda([v,deg(v)]))$.\\
    \hspace*{1em}\hspace*{1em} \textbf{for} $l^\prime$ from $deg(v)-1$ to $1$ \textbf{do}\\
    \hspace*{1em}\hspace*{1em}\hspace*{1em} $state([v,l^\prime]):=\delta(childState(l^\prime),state([v,l^\prime+1]),\lambda([v,l^\prime]))$.\\
    \hspace*{1em}\hspace*{1em} \textbf{end for}\\
    \hspace*{1em}\hspace*{1em} \textbf{if} $state([v,1]) \in F$ \textbf{then} $result(\varphi):=true$.\\
    \hspace*{1em}\hspace*{1em} \textbf{else} $result(\varphi):=false$.\\
    \hspace*{1em}\hspace*{1em} \textbf{end if}\\
    \hspace*{1em} \textbf{else}\\
    \hspace*{1em}\hspace*{1em} Let $l_0$ be the port with state ``parent''.\\
    \hspace*{1em}\hspace*{1em} \textbf{if} $l_0=1$ \textbf{then}\\
    \hspace*{1em}\hspace*{1em}\hspace*{1em} $state([v,deg(v)]):=\delta(childState(deg(v)),\lambda([v,deg(v)]))$.\\
    \hspace*{1em}\hspace*{1em}\hspace*{1em} \textbf{for} $l^\prime$ from $deg(v)-1$ to $2$ \textbf{do}\\
    \hspace*{1em}\hspace*{1em}\hspace*{1em}\hspace*{1em} $state([v,l^\prime]):=\delta(childState(l^\prime),state([v,l^\prime+1]),\lambda([v,l^\prime]))$.\\
    \hspace*{1em}\hspace*{1em}\hspace*{1em} \textbf{end for}\\
    \hspace*{1em}\hspace*{1em}\hspace*{1em} $state([v,1]):=\delta(state([v,2]),\lambda([v,1]))$.\\
    \hspace*{1em}\hspace*{1em} \textbf{else if} $l_0=deg(v)$ \textbf{then}\\
    \hspace*{1em}\hspace*{1em}\hspace*{1em} $state([v,1]):=\delta(childState(1),\lambda([v,1]))$.\\
    \hspace*{1em}\hspace*{1em}\hspace*{1em} \textbf{for} $l^\prime$ from $2$ to $deg(v)-1$ \textbf{do}\\
    \hspace*{1em}\hspace*{1em}\hspace*{1em}\hspace*{1em} $state([v,l^\prime]):=\delta(state([v,l^\prime-1]),childState(l^\prime),\lambda([v,l^\prime]))$.\\
    \hspace*{1em}\hspace*{1em}\hspace*{1em} \textbf{end for}\\
    \hspace*{1em}\hspace*{1em}\hspace*{1em} $state([v,deg(v)]):=\delta(state([v,deg(v)-1]),\lambda([v,deg(v)]))$.\\
    \hspace*{1em}\hspace*{1em} \textbf{else}\\
    \hspace*{1em}\hspace*{1em}\hspace*{1em} $state([v,1]):=\delta(childState(1),\lambda([v,1]))$.\\
    \hspace*{1em}\hspace*{1em}\hspace*{1em} $state([v,deg(v)]):=\delta(childState(deg(v)),\lambda([v,deg(v)]))$.\\
    \hspace*{1em}\hspace*{1em}\hspace*{1em} \textbf{for} $l^\prime$ from $2$ to $l_0-1$ \textbf{do}\\
    \hspace*{1em}\hspace*{1em}\hspace*{1em}\hspace*{1em} $state([v,l^\prime]):=\delta(state([v,l^\prime-1]),childState(l^\prime),\lambda([v,l^\prime]))$.\\
    \hspace*{1em}\hspace*{1em}\hspace*{1em} \textbf{end for}\\
    \hspace*{1em}\hspace*{1em}\hspace*{1em} \textbf{for} $l^\prime$ from $deg(v)-1$ to $l_0+1$ \textbf{do}\\
    \hspace*{1em}\hspace*{1em}\hspace*{1em}\hspace*{1em} $state([v,l^\prime]):=\delta(childState(l^\prime),state([v,l^\prime+1]),\lambda([v,l^\prime]))$.\\
    \hspace*{1em}\hspace*{1em}\hspace*{1em} \textbf{end for}\\
    \hspace*{1em}\hspace*{1em}\hspace*{1em} $state([v,l_0]):=\delta(state([v,l_0-1]),state([v,l_0+1]),\lambda([v,l_0]))$.\\
    \hspace*{1em}\hspace*{1em} \textbf{end if}\\
    \hspace*{1em}\hspace*{1em} $v$ sends message STATE($state([v,l_0])$) over port $l_0$.\\
    \hspace*{1em} \textbf{end if}\\
    \textbf{end if}\\
    \hline
    \end{tabular*}

\clearpage

\section{Distributed ordered tree decomposition of biconnected planar networks with bounded
diameter} \label{appendix-MSO-biconn-planar-diameter}

First, a breadth-first-search (BFS) tree rooted on the requesting
node is distributively constructed and stored in the network such
that each node $v$ stores the identifier of its parent in the
BFS-tree ($parent(v)$), and the states of the ports with respect to
the BFS-tree ($state(l)$ for each port $l$), which are either
``parent'', or ``child'', or ``non-tree'' \cite{BDLP08}.

Then, the requesting node sends messages to ask each node $v$ to get
the list of all its ancestors, denoted as $ancestor(v)$, and all its
neighbors ($neighbor(l)$ for each port $l$) within the complexity
bounds.

%

By a post-order traversal of the BFS-tree, $G$ can be triangulated
as follows. Each bidirectional link is seen as two arcs with reverse
directions. When traversing a node $v$, if all the arcs (starting
from $v$) $(v,w)$ have been traversed, then backtrack to the parent
of $v$ and traverse the next node; otherwise, for each arc $(v,w)$
that has not been traversed before, walk along the facial cycle
containing $(v,w)$ according to the cyclic ordering in each node and
$(v,w)$ is called the starting arc of this facial cycle. When all
the walks of those facial cycles returned to $v$, then backtrack to
the parent of $v$ and traverse the next node. When an arc
$(v^\prime,w^\prime)$ in a facial cycle is traversed, let
$(v_0,w_0)$ be the starting arc of the facial cycle, we imagine that
$v^\prime$ and $w^\prime$ are connected by virtual edges to $v_0$,
i.e. imagine $\{v_0,v^\prime,w^\prime\}$ as a triangle in the
triangulated graph, then the node $w^\prime$ stores locally the
starting arc $(v_0,w_0)$ and the information about the bag
corresponding to the triangle $\{v_0,v^\prime,w^\prime\}$.

After $G$ is triangulated, an ordered tree decomposition can be
obtained by selecting some bag stored in the requesting node as the
root bag, and connecting together all the bags (corresponding to the
triangles) depending on whether they share a non-BFS-tree link or
not.

In the following, we describe the distributed algorithm in detail by
giving the pseudo-code for the message processing at each node $v$.

\bigskip \noindent
  \begin{tabular*}{\textwidth}{l@{\extracolsep{\fill}}}
    \hline
    Initialization\\
    \hline
    The requesting node sends messages to ask each node to\\
    \hspace*{1em} get the list of its ancestors in the BFS-tree and all its neighbors.\\
    For each node $v$, let $ancestor(v)$ be the list of its ancestors.\\
    For each port $l$, let $neighbor(l)$ be the neighbor connected to $l$.\\
    The requesting node sets $traversed(1):=true$, and sends POSTTRAVERSE over port $1$.\\
    \hline
    \\
    \hline
    Message POSTTRAVERSE over port $l$\\
    \hline
    \textbf{if} $v$ is a leaf in the BFS-tree \textbf{then}\\
    \hspace*{1em} \textbf{if} there exist $l^\prime$ such that $arcVisited(v,neighbor(l^\prime))=false$ \textbf{then}\\
    \hspace*{1em}\hspace*{1em} \textbf{for each} $l^\prime: arcVisited(v,neighbor(l^\prime))=false$ \textbf{do}\\
    \hspace*{1em}\hspace*{1em}\hspace*{1em} $arcVisited(v,neighbor(l^\prime)):=true$.\\
    \hspace*{1em}\hspace*{1em}\hspace*{1em} $v$ sends FACESTART(($v$,$neighbor(l^\prime)$), $ancestor(v)$, $ancestor(v)$) over $l^\prime$.\\
    \hspace*{1em}\hspace*{1em} \textbf{end for}\\
    \hspace*{1em} \textbf{else}\\
    \hspace*{1em}\hspace*{1em} $v$ sends BACKTRACK over the port $l^\prime$ such that $state(l^\prime)=$``parent''.\\
    \hspace*{1em} \textbf{end if}\\
    \textbf{else}\\
    \hspace*{1em} Let $l^\prime$ be the minimal port such that $state(l^\prime)=$``child''.\\
    \hspace*{1em} $v$ sets $traversed(l^\prime):=true$ and sends message POSTTRAVERSE over $l^\prime$.\\
    \textbf{end if}\\
    \hline
  \end{tabular*}

\clearpage \noindent
  \begin{tabular*}{\textwidth}{l@{\extracolsep{\fill}}}
    \\
    \hline
    Message FACESTART(($u_1$,$u_2$), $(w_1,\cdots,w_r)$, $(w^\prime_1,\cdots,w^\prime_s)$) over the port $l$\\
    \hline
    $arcVisited(neighbor(l),v):=true$.\\
    Let $l^\prime$ be the port such that $(v,neighbor(l^\prime))$ is immediately before $(v,neighbor(l))$ in
the cyclic ordering.\\
    \textbf{if} $v=u_2$ \textbf{then}\\
    \hspace*{1em} $v$ sends message FACESTART(($u_1$,$u_2$),$(w_1,\cdots,w_r)$, $ancestor(v)$) over port $l^\prime$.\\
    \textbf{else}\\
    \hspace*{1em} $Bag(u_1,neighbor(l),v):=\langle (u_1,u_2),list_{3k+1}((w_1,\cdots,w_r) \cdot (w^\prime_1,\cdots,w^\prime_s) \cdot
    ancestor(v))\rangle$.\\
    \hspace*{1em} \% \textsf{$list_{3k+1}(w_1,\cdots,w_p)$ ($p \le 3k+1$) generates a list of length $3k+1$} \\
    \hspace*{1em}\hspace*{1em} \textsf{from $w_1,\cdots,w_p$ by repeating $w_p$ after $w_1,\cdots,w_p$.}\\
    \hspace*{1em} $v$ sends message ACKFACESTART(($u_1$,$u_2$), $(w_1,\cdots,w_r)$, $ancestor(v)$) over port $l$.\\
    \hspace*{1em} \textbf{if} $neighbor(l^\prime)=u_1$ \textbf{then} \hspace*{1em}\% \textsf{$\{u_1,u_2,v\}$ is a triangle.}\\
    \hspace*{1em}\hspace*{1em} $v$ sends message FACEOVER(($u_1$,$u_2$), $neighbor(l)$, $(w^\prime_1,\cdots,w^\prime_s)$, $ancestor(v)$) over port $l^\prime$.\\
    \hspace*{1em} \textbf{else}\\
    \hspace*{1em}\hspace*{1em} $v$ sends message FACEWALK(($u_1$,$u_2$), $(w_1,\cdots,w_r)$, $ancestor(v)$) over port $l^\prime$.\\
    \hspace*{1em} \textbf{end if}\\
    \hspace*{1em} $arcVisited(v,neighbor(l^\prime)):=true$.\\
    \textbf{end if}\\
    \hline
    \\
    \hline
    Message ACKFACESTART(($u_1$,$u_2$), $(w_1,\cdots,w_r)$, $(w^\prime_1,\cdots,w^\prime_s)$) over port $l$\\
    \hline
    $Bag(u_1,v, neighbor(l)):=\langle (u_1,u_2),list_{3k+1}((w_1,\cdots,w_r) \cdot ancestor(v) \cdot (w^\prime_1,\cdots,w^\prime_s))\rangle$.\\
    \hline
    \\
    \hline
    Message FACEOVER(($u_1$,$u_2$), $v^\prime$, $(w_1,\cdots,w_r)$, $(w^\prime_1,\cdots,w^\prime_s)$) over port $l$\\
    \hline
    $arcVisited(neighbor(l),v):=true$.\\
    $Bag(v,v^\prime, neighbor(l)):=\langle (u_1,u_2),list_{3k+1}(ancestor(v) \cdot (w_1,\cdots,w_r) \cdot (w^\prime_1,\cdots,w^\prime_s))\rangle$.\\
    \textbf{if} $arcVisited(neighbor(l^\prime),v)=true$ for each $l^\prime$ \textbf{then}\\
    \hspace*{1em} \textbf{if} $v$ is the requesting node \textbf{then}\\
    \hspace*{1em}\hspace*{1em} $root:=(v,v^{\prime},neighbor(l))$, $bLeafBag(v,v^{\prime},neighbor(l)):=false$.\\
    \hspace*{1em}\hspace*{1em} $v$ sends message ANTIBAGVISIT($(v,v^{\prime},neighbor(l))$,$Bag(v,v^{\prime},neighbor(l))$) over port $l$.\\
    \hspace*{1em} \textbf{else}\\
    \hspace*{1em}\hspace*{1em} $v$ sends message BACKTRACK over link $l^\prime$ such that $state(l^\prime)=$``parent''.\\
    \hspace*{1em} \textbf{end if}\\
    \textbf{end if}\\
    \hline
    \\
    \hline
    Message FACEWALK(($u_1$,$u_2$), $(w_1,\cdots,w_r)$, $(w^\prime_1,\cdots,w^\prime_s)$) over port $l$\\
    \hline
    $arcVisited(neighbor(l),v):=true$.\\
    $Bag(u_1,neighbor(l),v):=\langle (u_1,u_2),list_{3k+1}((w_1,\cdots,w_r) \cdot (w^\prime_1,\cdots,w^\prime_s) \cdot
    ancestor(v))\rangle$.\\
    Let $l^\prime$ be the port such that\\
    \hspace*{1em} $(v,neighbor(l^\prime))$ is immediately before $(v,neighbor(l))$ in the cyclic
    ordering.\\
    \textbf{if} $neighbor(l^\prime)=u_1$ \textbf{then}\\
    \hspace*{1em} $v$ sends message FACEOVER(($u_1$,$u_2$),$(w^\prime_1,\cdots,w^\prime_s)$, $ancestor(v)$) over port $l^\prime$.\\
    \textbf{else}\\
    \hspace*{1em} $v$ sends message FACEWALK(($u_1$,$u_2$),$(w_1,\cdots,w_r)$, $ancestor(v)$) over port $l^\prime$.\\
    \textbf{end if}\\
    $arcVisited(v,neighbor(l^\prime)):=true$.\\
    \hline
    \end{tabular*}

\noindent
  \begin{tabular*}{\textwidth}{l@{\extracolsep{\fill}}}
    \hline
    Message BACKTRACK over port $l$\\
    \hline
    \textbf{if} there exists $l^\prime$ such that
    $state(l^\prime)=$``child'' and $traversed(l^\prime)=false$
    \textbf{then}\\
    \hspace*{1em} Let $l^\prime$ be the minimal port such that $state(l^\prime)=$``child'' and $traversed(l^\prime)=false$.\\
    \hspace*{1em} $v$ sets $traversed(l^\prime):=true$ and sends POSTTRAVERSE over $l^\prime$.\\
    \textbf{else if} there exists $l^\prime$ such that $arcVisited(v,neighbor(l^\prime))=false$ \textbf{then}\\
    \hspace*{1em} \textbf{for each} port $l^\prime$ such that $arcVisited(v,neighbor(l^\prime))=false$ \textbf{do}\\
    \hspace*{1em}\hspace*{1em} $arcVisited(v,neighbor(l^\prime)):=true$.\\
    \hspace*{1em}\hspace*{1em} $v$ sends FACESTART(($v$,$neighbor(l^\prime)$),$ancestor(v)$, $ancestor(v)$) over port $l^\prime$. \\
    \hspace*{1em} \textbf{end for}\\
    \textbf{else}\\
    \% \textsf{All the children of $v$ have been traversed and all the
    facial cycles containing $v$ have been visited}.\\
    \hspace*{1em} \textbf{if} $v$ is not the requesting node \textbf{then}\\
    \hspace*{1em}\hspace*{1em} $v$ sends message BACKTRACK over the port $l^\prime$ such that $state(l^\prime)=$``parent''.\\
    \hspace*{1em} \textbf{else}\\
    \hspace*{1em}\hspace*{1em} $v$ selects some bag $(v^\prime,v^{\prime\prime},v^{\prime\prime\prime})$ stored in it.\\
    \hspace*{1em}\hspace*{1em} $root:=(v^\prime,v^{\prime\prime},v^{\prime\prime\prime})$, $bLeafBag(v^\prime,v^{\prime\prime},v^{\prime\prime\prime}):=false$.\\
    \hspace*{1em}\hspace*{1em} \textbf{if} $v=v^{\prime}$ \textbf{then}\\
    \hspace*{1em}\hspace*{1em} \% \textsf{$(v^{\prime\prime\prime},v)$ is the last arc of the facial
    cycle.}\\
    \hspace*{1em}\hspace*{1em}\hspace*{1em} Let $l^\prime$ be the port such that $neighbor(l^\prime)=v^{\prime\prime\prime}$.\\
    \hspace*{1em}\hspace*{1em}\hspace*{1em} $v$ sends message ANTIBAGVISIT($(v,v^{\prime\prime},v^{\prime\prime\prime})$,
    $Bag(v,v^{\prime\prime},v^{\prime\prime\prime})$) over port $l^\prime$.\\
    \hspace*{1em}\hspace*{1em}\hspace*{1em} \textbf{if} $state(l^\prime)=$``non-tree'' \textbf{then}\\
    \hspace*{1em}\hspace*{1em}\hspace*{1em} \% \textsf{$(v^{\prime\prime\prime},v)$ is the last arc of the facial
    cycle.}\\
    \hspace*{1em}\hspace*{1em}\hspace*{1em}\hspace*{1em} $v$ sends message NEWFACEBAG($(v,v^{\prime\prime},v^{\prime\prime\prime})$,
    $Bag(v,v^{\prime\prime},v^{\prime\prime\prime})$) over port $l^\prime$.\\
    \hspace*{1em}\hspace*{1em}\hspace*{1em} \textbf{end if}\\
    \hspace*{1em}\hspace*{1em} \textbf{else if} $v=v^{\prime\prime}$ \textbf{then}\\
    \hspace*{1em}\hspace*{1em} \% \textsf{$(v,v^{\prime\prime})$ is the starting arc of the facial
    cycle.}\\
    \hspace*{1em}\hspace*{1em}\hspace*{1em} Let $l^\prime$ be the port such that $neighbor(l^\prime)=v^{\prime}$.\\
    \hspace*{1em}\hspace*{1em}\hspace*{1em} Let $l^{\prime\prime}$ be the port such that $neighbor(l^{\prime\prime})=v^{\prime\prime\prime}$.\\
    \hspace*{1em}\hspace*{1em}\hspace*{1em} $v$ sends message BAGVISIT($(v^\prime,v,v^{\prime\prime\prime})$,
    $Bag(v^\prime,v,v^{\prime\prime\prime})$) over port $l^{\prime\prime}$.\\
    \hspace*{1em}\hspace*{1em}\hspace*{1em} \textbf{if} $state(l^\prime)=$``non-tree'' \textbf{then}\\
    \hspace*{1em}\hspace*{1em}\hspace*{1em}\hspace*{1em} $v$ sends message NEWFACEBAG($(v^\prime,v,v^{\prime\prime\prime})$,
    $Bag(v^\prime,v,v^{\prime\prime\prime})$) over port $l^\prime$.\\
    \hspace*{1em}\hspace*{1em}\hspace*{1em} \textbf{end if}\\
    \hspace*{1em}\hspace*{1em} \textbf{else}\\
    \hspace*{1em}\hspace*{1em} \% $v=v^{\prime\prime\prime}$.\\
    \hspace*{1em}\hspace*{1em}\hspace*{1em} Let $l^\prime$ be the port such that $neighbor(l^\prime)=v^{\prime\prime}$.\\
    \hspace*{1em}\hspace*{1em}\hspace*{1em} Let $l^{\prime\prime}$ be the port such that $(v,neighbor(l^{\prime\prime}))$ is immediately before $(v,v^{\prime\prime})$ in the cyclic ordering.\\
    \hspace*{1em}\hspace*{1em}\hspace*{1em} $v$ sends message ANTIBAGVISIT($(v^\prime,v^{\prime\prime},v)$,
    $Bag(v^\prime,v^{\prime\prime},v)$) over port $l^\prime$.\\
    \hspace*{1em}\hspace*{1em}\hspace*{1em} \textbf{if} $state(l^\prime)=$``non-tree'' \textbf{then}\\
    \hspace*{1em}\hspace*{1em}\hspace*{1em}\hspace*{1em} $v$ sends message NEWFACEBAG($(v^\prime,v^{\prime\prime},v)$,
    $Bag(v^\prime,v^{\prime\prime},v)$) over port $l^\prime$.\\
    \hspace*{1em}\hspace*{1em}\hspace*{1em} \textbf{end if}\\
    \hspace*{1em}\hspace*{1em}\hspace*{1em} $v$ sends message BAGVISIT($(v^\prime,v^{\prime\prime},v)$,
    $Bag(v^\prime,v^{\prime\prime},v)$) over port $l^{\prime\prime}$.\\
    \hspace*{1em}\hspace*{1em} \textbf{end if}\\
    \hspace*{1em} \textbf{end if}\\
    \textbf{end if}\\
    \hline
  \end{tabular*}

\noindent
  \begin{tabular*}{\textwidth}{l@{\extracolsep{\fill}}}
    \hline
    Message BAGVISIT($bagId$, $bag$) over port $l$\\
    \hline
    Let $bagId:=(w_1,w_2,w_3)$, $bag:=\langle (u_1,u_2),(w^\prime_1,\cdots,w^\prime_{3k+1}) \rangle$.\\
    \textbf{if} $v=u_1$ \textbf{then}\\
    \hspace*{1em} Let $(v,v^{\prime},v^{\prime\prime})$ be the stored bag such that
    $Bag(v,v^{\prime},v^{\prime\prime})=\langle (u_1,u_2),(\cdots) \rangle$.\\
    \hspace*{1em} $father(v,v^{\prime},v^{\prime\prime}):=(bagId,bag)$.\\
    \hspace*{1em} \textbf{if} $state(l)=$``non-tree'' \textbf{then}\\
    \hspace*{1em}\hspace*{1em} $v$ sends message NEWFACEBAG$((v,v^{\prime},v^{\prime\prime}), Bag(v,v^{\prime},v^{\prime\prime}))$ over port $l$.\\
    \hspace*{1em} \textbf{else} $bLeafBag(v,v^{\prime},v^{\prime\prime}):=true$.\\
    \hspace*{1em} \textbf{end if}\\
    \textbf{else}\\
    \hspace*{1em} $father(u_1,neighbor(l),v):=(bagId,bag)$, $bLeafBag(u_1,neighbor(l),v):=false$.\\
    \hspace*{1em} \textbf{if} $state(l)=$``non-tree'' \textbf{then}\\
    \hspace*{1em}\hspace*{1em} $v$ sends message NEWFACEBAG$((u_1,neighbor(l),v), Bag(u_1,neighbor(l),v))$ over port $l$.\\
    \hspace*{1em} \textbf{end if}\\
    \hspace*{1em} Let $l^\prime$ be the port such that $(v,neighbor(l^\prime))$ is immediately before $(v,neighbor(l))$ in the cyclic ordering.\\
    \hspace*{1em} $v$ sends message BAGVISIT($(u_1,neighbor(l),v)$,$Bag(u_1,neighbor(l),v)$) over port $l^\prime$.\\
    \textbf{end if}\\
    \hline
    \\
    \hline
    Message ANTIBAGVISIT($bagId$, $bag$) over port $l$\\
    \hline
    Let $bagId:=(w_1,w_2,w_3)$, $bag:=\langle(u_1,u_2),(w^\prime_1,\cdots,w^\prime_{3k+1})\rangle$.\\
    Let $l^\prime$ be the port such that $(v,neighbor(l))$ is immediately before
    $(v,neighbor(l^\prime))$ in the cyclic ordering.\\
    \textbf{if} $neighbor(l^\prime) = u_1$ \textbf{then}\\
    \hspace*{1em} \textbf{if} $state(l^\prime)=$``non-tree'' \textbf{then}\\
    \hspace*{1em}\hspace*{1em}
    $bLeafBag(u_1,v,neighbor(l)):=false$.\\
    \hspace*{1em}\hspace*{1em} $v$ sends NEWFACEBAG($(u_1,v,neighbor(l)), Bag(u_1,v,neighbor(l))$) over $l^\prime$.\\
    \hspace*{1em} \textbf{else} $bLeafBag(u_1,v,neighbor(l)):=true$.\\
    \hspace*{1em} \textbf{end if}\\
    \textbf{else}\\
    \hspace*{1em} $father(u_1,neighbor(l^\prime),v):=(bagId,bag)$, $bLeafBag(u_1,neighbor(l^\prime),v):=false$.\\
    \hspace*{1em} $v$ sends message ANTIBAGVISIT$((u_1,neighbor(l^\prime),v),Bag(u_1,neighbor(l^\prime),v))$ over
    $l^\prime$.\\
    \hspace*{1em} \textbf{if} $state(l^\prime)=$``non-tree'' \textbf{then}\\
    \hspace*{1em}\hspace*{1em} $v$ sends message NEWFACEBAG$((u_1,neighbor(l^\prime),v),Bag(u_1,neighbor(l^\prime),v))$ over port $l^\prime$.\\
    \hspace*{1em} \textbf{end if}\\
    \textbf{end if}\\
    \hline
    \\
    \hline
    Message NEWFACEBAG($bagId$,$bag$) over port $l$\\
    \hline
    Let $(v^\prime,v^{\prime\prime},v^{\prime\prime\prime})$ be the stored bag such
    that $(neighbor(l),v)=(v^\prime,v^{\prime\prime})$ or $(v^{\prime\prime},v^{\prime\prime\prime})$ or $(v^{\prime\prime\prime},v^\prime)$.\\
    \% \textsf{$(neighbor(l),v)=(v^\prime,v^{\prime\prime})$: $(neighbor(l),v)$ is the first arc traversed in the new facial
    cycle.}\\
    \% \textsf{$(neighbor(l),v)=(v^{\prime\prime\prime},v^\prime)$: $(neighbor(l),v)$ is the last arc traversed in the new facial
    cycle.}\\
    $father(v^\prime,v^{\prime\prime},v^{\prime\prime\prime}):=(bagId,bag)$,
    $bLeafBag(v^\prime,v^{\prime\prime},v^{\prime\prime\prime}):=false$.\\
    \textbf{if} $(neighbor(l),v)=(v^\prime,v^{\prime\prime})$ \textbf{then}\\
    \hspace*{1em} Let $l^\prime$ be the port such that $neighbor(l^\prime)=v^{\prime\prime\prime}$.\\
    \hspace*{1em} $v$ sends message BAGVISIT($(v^\prime,v,v^{\prime\prime\prime})$,
    $Bag(v^\prime,v,v^{\prime\prime\prime})$) over port $l^{\prime}$.\\
    \textbf{else if} $(neighbor(l),v)=(v^{\prime\prime\prime},v^\prime)$ \textbf{then} \\
    \hspace*{1em} $v$ sends message ANTIBAGVISIT($(v,v^{\prime\prime},v^{\prime\prime\prime})$,
    $Bag(v,v^{\prime\prime},v^{\prime\prime\prime})$) over port $l$.\\
    \textbf{else}\\
    \hspace*{1em} Let $l^\prime$ be the port such that $(v,neighbor(l^\prime))$ is immediately before $(v,neighbor(l))$ in the cyclic ordering.\\
    \hspace*{1em} $v$ sends message ANTIBAGVISIT($(v^\prime,v^{\prime\prime},v)$,
    $Bag(v^\prime,v^{\prime\prime},v)$) over port $l$.\\
    \hspace*{1em} $v$ sends message BAGVISIT($(v^\prime,v^{\prime\prime},v)$,
    $Bag(v^\prime,v^{\prime\prime},v)$) over port $l^{\prime}$.\\
    \textbf{end if}\\
    \hline
    \end{tabular*}

\clearpage

\section{Distributed ordered tree decomposition of general planar networks with bounded
diameter}\label{appendix-MSO-gene-planar-diameter}

\vspace*{-2mm}

First, a breadth-first-search (BFS) tree rooted on the requesting
node is distributively constructed and stored in the network such
that each node $v$ stores the identifier of its parent in the
BFS-tree ($parent(v)$), and the states of the ports with respect to
the BFS-tree ($state(l)$ for each port $l$), which are either
``parent'', or ``child'', or ``non-tree'' \cite{BDLP08}.

Then, a distributed depth-first-search can be done to decompose the
planar network with bounded diameter into biconnected components
(also called blocks) \cite{Turau06}. An ordered tree decomposition
for each block is constructed. Finally these ordered tree
decompositions are connected together to get the complete tree
decomposition of the whole network.

We first describe the distributed algorithm to decompose the network
into blocks by giving the pseudo-code for the message processing at
each node $v$.

\medskip \noindent
%
  \begin{tabular*}{\textwidth}{l@{\extracolsep{\fill}}}
    \hline
    Initialization\\
    \hline
    The requesting node $v_0$ sets $DFSDepth(v_0):=DFSLow(v_0):=0$, $DFSVisited(v_0):=false$.\\
    The requesting node $v_0$ sets $DFSState(l):=$``unvisited'' for each port $l$.\\
    The requesting node sets $DFSState(1):=$``child'' and sends message FORWARD($1$,$0$) over port $1$.\\
    \hline
    \\
    \hline
    Message FORWARD($nextBlockId$, $parentDepth$) over port $l$\\
    \hline
    \textbf{if} $DFSVisited(v)=false$ \textbf{then}\\
    \hspace*{1em} $DFSVisited(v):=true$, $DFSState(l):=$``parent'', $DFSState(l^\prime):=$``unvisited'' for each port $l^\prime \ne l$.\\
    \hspace*{1em} $DFSDepth(v):=parentDepth+1$, $DFSLow(v):=DFSDepth(v)$.\\
    \hspace*{1em} \textbf{if} $v$ has at least two ports \textbf{then}\\
    \hspace*{1em}\hspace*{1em} Let $l^\prime$ be the minimal port such that $l^\prime \ne l$.\\
    \hspace*{1em}\hspace*{1em} $DFSState(l^\prime):=$``child'', $v$ sends message FORWARD($nextBlockId$, $DFSDepth(v)$) over $l^\prime$.\\
    \hspace*{1em} \textbf{else} $v$ sends message BACKTRACK($nextBlockId$,$DFSLow(v)$) and message BLOCKACK over $l$. \\
    \hspace*{1em} \textbf{end if}\\
    \textbf{else} $DFSState(l):=$``non-tree-forward'', $v$ sends message RESTART$(nextBlockId,DFSDepth(v))$ over $l$.\\
    \textbf{end if}\\
    \hline
    \\
    \hline
    Message BACKTRACK($nextBlockId$, $childLow$) over port $l$\\
    \hline
    \textbf{if} $childLow > = DFSDepth(v)$ \textbf{then}\\
    \hspace*{1em} $articulation(v):=true$.\\
    \hspace*{1em} \textbf{if} $childLow = DFSDepth(v)$ \textbf{then}\\
    \hspace*{1em}\hspace*{1em} $DFSState(l):=$``closed'', $blockIds:=blockIds \cup \{nextBlockId\}$, $blockPorts(nextBlockId):=\{l\}$.\\
    \hspace*{1em}\hspace*{1em} $v$ sends message BLOCKINFORM$(nextBlockId)$ over $l$.\\
    \hspace*{1em} \textbf{else if} $childLow > DFSDepth(v)$ \textbf{then} \\
    \hspace*{1em}\hspace*{1em} $DFSState(l):=$``childBridge'', $v$ sends message BRIDGEINFORM over $l$.\\
    \hspace*{1em} \textbf{end if}\\
    \textbf{else} $DFSState(l):=$``backtracked'', $DFSLow(v):=\min\{DFSLow(v),childLow\}$.\\
    \textbf{end if}\\
    \textbf{if} there exist at least one port $l^\prime$ such that $DFSState(l^\prime)=$``unvisited'' \textbf{then}\\
    \hspace*{1em} Let $l^\prime$ be the minimal port such that $DFSState(l^\prime)=$``unvisited''.\\
    \hspace*{1em} $DFSState(l^\prime):=$``child''.\\
    \hspace*{1em} \textbf{if} $childLow=DFSDepth(v)$ \textbf{then}\\
    \hspace*{1em}\hspace*{1em} $v$ sends message FORWARD($nextBlockId+1$, $DFSDepth(v)$) over $l^\prime$.\\
    \hspace*{1em} \textbf{else} $v$ sends message FORWARD($nextBlockId$, $DFSDepth(v)$) over $l^\prime$.\\
    \hspace*{1em} \textbf{end if}\\
    \textbf{else}\\
    \hspace*{1em} \textbf{if} $v$ is not the requesting node \textbf{then}\\
    \hspace*{1em}\hspace*{1em} $v$ sends message BACKTRACK($nextBlockId$,$DFSLow(v)$) over $l^\prime$ such that $DFSState(l^\prime)=$``parent''.\\
    \hspace*{1em} \textbf{end if}\\
    \textbf{end if}\\
    \hline
  \end{tabular*}

\clearpage

  \begin{tabular*}{\textwidth}{l@{\extracolsep{\fill}}}
    \hline
    Message RESTART($nextBlockId$, $ancestorDepth$) over port $l$.\\
    \hline
    $DFSState(l)$:=``non-tree-backward'', $DFSLow(v):=\min\{DFSLow(v),ancestorDepth\}$. \\
    \textbf{if} there exist at least one port $l^\prime$ such that $DFSState(l^\prime)=$``unvisited'' \textbf{then}\\
    \hspace*{1em} Let $l^\prime$ be the minimal port such that $DFSState(l^\prime)=$``unvisited''.\\
    \hspace*{1em} $DFSState(l^\prime):=$``child''.\\
    \hspace*{1em} $v$ sends message FORWARD($nextBlockId$,$DFSDepth(v)$) over $l^\prime$.\\
    \textbf{else}\\
    \hspace*{1em} \textbf{if} $v$ is not the requesting node \textbf{then}\\
    \hspace*{1em}\hspace*{1em} $v$ sends message BACKTRACK($nextBlockId$,$DFSLow(v)$) over $l^\prime$ such that $DFSState(l^\prime)=$``parent''.\\
    \hspace*{1em}\hspace*{1em} \textbf{if} there are no ports $l^\prime$ such that \\
    \hspace*{1em}\hspace*{1em}\hspace*{1em} $DFSState(l^\prime)=$``closed'' or ``backtracked'' or ``childBridge'' \textbf{then}\\
    \hspace*{1em}\hspace*{1em}\hspace*{1em} \% \textsf{$v$ has no children in the DFS-tree.}\\
    \hspace*{1em}\hspace*{1em}\hspace*{1em} $v$ sends message BLOCKACK over the port $l^\prime$ such that $DFSState(l^\prime)=$``parent''. \\
    \hspace*{1em}\hspace*{1em} \textbf{end if}\\
    \hspace*{1em} \textbf{end if}\\
    \textbf{end if}\\
    \hline
    \\
    \hline
    Message BLOCKINFORM($blockId$) over port $l$.\\
    \hline
    \textbf{if} $blockId \not \in blockIds$ \textbf{then}\\
    \hspace*{1em} $blockIds:=blockIds \cup \{blockId\}$.\\
    \hspace*{1em} $blockPorts(blockId):= \{l^\prime | DFSState(l^\prime)=\mbox{``parent'' or ``non-tree-backward'' or ``backtracked''}\}$.\\
    \hspace*{1em} $v$ sends message BLOCKPORT$(blockId)$ over all $l^\prime$ such that $DFSState(l^\prime)=$``non-tree-backward''.\\
    \hspace*{1em} \textbf{if} there exists at least one port $l^\prime$ such that $DFSState(l^\prime)=$``backtracked'' \textbf{then}\\
    \hspace*{1em}\hspace*{1em} $v$ sends message BLOCKINFORM$(blockId)$ over all ports $l^\prime$ such that $DFSState(l^\prime)=$``backtracked''.\\
    \hspace*{1em} \textbf{else}\\
    \hspace*{1em}\hspace*{1em} $v$ sends message INFORMOVER$(blockId)$ over $l^\prime$ such that $DFSState(l^\prime)=$``parent''.\\
    \hspace*{1em}\hspace*{1em} \textbf{if} there are no ports $l^\prime$ such that $DFSState(l^\prime)=$``closed'' or ``backtracked'' or ``childBridge'' \textbf{then}\\
    \hspace*{1em}\hspace*{1em}\hspace*{1em} $v$ sends message BLOCKACK over $l^\prime$ such that $DFSState(l^\prime)=$``parent''. \\
    \hspace*{1em}\hspace*{1em} \textbf{end if}\\
    \hspace*{1em} \textbf{end if}\\
    \textbf{end if}\\
    \hline
    \\
    \hline
    Message BRIDGEINFORM over port $l$.\\
    \hline
    $DFSState(l):=$``parentBridge''.\\
    \hline
    \\
    \hline
    Message BLOCKPORT($blockId$) over port $l$.\\
    \hline
    $blockPorts(blockId):=blockPorts(blockId) \cup \{l\}$.\\
    \hline
  \end{tabular*}

  \begin{tabular*}{\textwidth}{l@{\extracolsep{\fill}}}
    \hline
    Message INFORMOVER($blockId$) over port $l$.\\
    \hline
    $bInformOver(l):=true$.\\
    \textbf{if} $DFSState(l) = $``backtracked'' \textbf{then}\\
    \hspace*{1em} \textbf{if} $bInformOver(l^\prime)=true$ for each $l^\prime$ such that $DFSState(l^\prime)=$``backtracked'' \textbf{then}\\
    \hspace*{1em}\hspace*{1em} $v$ sends message INFORMOVER($blockId$) over $l^\prime$ such that $DFSState(l^\prime)=$``parent''.\\
    \hspace*{1em} \textbf{end if}\\
    \textbf{else}\hspace*{1em} \% \textsf{$DFSState(l) = $``closed''.}\\
    \hspace*{1em} \textbf{if} $bInformOver(l^\prime)=true$ for each port $l^\prime$ such that $DFSState(l^\prime)=$``closed'', \\
    \hspace*{1em}\hspace*{1em} and $bBlockAck(l^\prime)=true$ for each port $l^\prime$ \\
    \hspace*{1em}\hspace*{1em}\hspace*{1em} such that $DFSState(l^\prime)=$``closed'' or ``backtracked'' or ``childBridge'' \textbf{then}\\
    \hspace*{1em}\hspace*{1em} \textbf{if} $v$ is not the requesting node \textbf{then}\\
    \hspace*{1em}\hspace*{1em}\hspace*{1em} $v$ sends BLOCKACK over the port $l^\prime$ such that $DFSState(l^\prime)=$``parent'' or ``parentBridge''.\\
    \hspace*{1em}\hspace*{1em} \textbf{else}\\
    \hspace*{1em}\hspace*{1em}\hspace*{1em} Let $T$ be the distributively stored BFS-tree.\\
    \hspace*{1em}\hspace*{1em}\hspace*{1em} $v$ sends message STARTDECOMP over all $l^\prime$ such that $DFSState(l^\prime)=$``closed'' or ``childBridge''.\\
    \hspace*{1em}\hspace*{1em}\hspace*{1em} \textbf{for each} $id \in blockIds$ such that there is $l^\prime$ \\
    \hspace*{1em}\hspace*{1em}\hspace*{1em}\hspace*{1em} satisfying $l^\prime \in blockPorts(id)$ and $DFSState(l^\prime)=$``closed'' \textbf{do}\\
    \hspace*{1em}\hspace*{1em}\hspace*{1em}\hspace*{1em} $v$ starts the tree decomposition for the block $(V_{id},E_{id})$\\
    \hspace*{1em}\hspace*{1em}\hspace*{1em}\hspace*{1em} by using $T[V_{id}]$, the subgraph of $T$ induced by $V_{id}$.\\
    \hspace*{1em}\hspace*{1em}\hspace*{1em} \textbf{end for}\\
    \hspace*{1em}\hspace*{1em} \textbf{end if}\\
    \hspace*{1em} \textbf{end if}\\
    \textbf{end if}\\
    \hline
    \\
    \hline
    Message BLOCKACK over port $l$.\\
    \hline
    $bBlockAck(l):=true$.\\
    \textbf{if} $bInformOver(l^\prime)=true$ for each port $l^\prime$ such that $DFSState(l^\prime)=$``closed'', \\
    \hspace*{1em} and $bBlockAck(l^\prime)=true$ for each port $l^\prime$ \\
    \hspace*{1em}\hspace*{1em} such that $DFSState(l^\prime)=$``closed'' or ``backtracked'' or ``childBridge'' \textbf{then}\\
    \hspace*{1em} \textbf{if} $v$ is not the requesting node \textbf{then}\\
    \hspace*{1em}\hspace*{1em} $v$ sends message BLOCKACK over the port $l^\prime$ \\
    \hspace*{1em}\hspace*{1em}\hspace*{1em} such that $DFSState(l^\prime)=$``parent'' or ``parentBridge''.\\
    \hspace*{1em} \textbf{else}\\
    \hspace*{1em}\hspace*{1em} Let $T$ be the distributively stored BFS-tree.\\
    \hspace*{1em}\hspace*{1em} $v$ sends message STARTDECOMP over all ports $l^\prime$ \\
    \hspace*{1em}\hspace*{1em}\hspace*{1em} such that $DFSState(l^\prime)=$``closed'' or ``childBridge''.\\
    \hspace*{1em}\hspace*{1em} \textbf{for each} $id \in blockIds$ such that there is $l^\prime$ \\
    \hspace*{1em}\hspace*{1em}\hspace*{1em} satisfying $l^\prime \in blockPorts(id)$ and $DFSState(l^\prime)=$``closed'' \textbf{do}\\
    \hspace*{1em}\hspace*{1em}\hspace*{1em} $v$ starts the tree decomposition for the block $(V_{id},E_{id})$\\
    \hspace*{1em}\hspace*{1em}\hspace*{1em} by using $T[V_{id}]$, the subgraph of $T$ induced by $V_{id}$.\\
    \hspace*{1em}\hspace*{1em} \textbf{end for}\\
    \hspace*{1em} \textbf{end if}\\
    \textbf{end if}\\
    \hline
    \\
    \hline
    Message STARTDECOMP over port $l$.\\
    \hline
    $v$ sends message STARTDECOMP over all ports $l^\prime$ such that \\
    \hspace*{1em} $DFSState(l^\prime)=$``closed'' or ``backtracked'' or
    ``childBridge''.\\
    \textbf{for each} $id \in blockIds$ such that there is $l^\prime$ satisfying $l^\prime \in blockPorts(id)$ and $DFSState(l^\prime)=$``closed'' \textbf{do}\\
    \hspace*{1em} Let $T$ be the distributively stored BFS-tree.\\
    \hspace*{1em} $v$ starts the tree decomposition for the block $(V_{id},E_{id})$ by using $T[V_{id}]$, the subgraph of $T$ induced by $V_{id}$.\\
    \textbf{end for}\\
    \hline
  \end{tabular*}
\\
\\
\\
Suppose now an ordered tree decomposition of each nontrivial block
of the network has been constructed, we show how to connect them
together to get a complete ordered tree decomposition with width
$3k+1$ and rank $2$ of the whole network.
\begin{enumerate}
\item At first each node $v$ is replaced by the set of virtual nodes
\[\{[v,l^\prime]|
    DFSState(l^\prime)=\mbox{``closed'' or ``childBridge'' or ``parentBridge'' or ``parent''}\}.\]

The intuition of the virtual nodes for $v$ is to have one virtual
node for each block to which it belongs.

\item The ordered bag for each virtual node $[v,l^\prime]$ is $list_{3k+1}((v))$, a list of length $3k+1$ with $v$ at each position.

\item The ordered bag corresponding to the virtual nodes are connected to the bags in the ordered tree decomposition of blocks as follows.
\begin{itemize}
\item The ordered bag corresponding to $[v,l^\prime]$ such that $DFSState(l^\prime)=$``closed'' or
``parent'' is connected to an ordered bag stored in $v$ in the
ordered tree decomposition of the block $id \in blockIds$ such that
$l^\prime \in blockPorts(id)$.
\item The ordered bag corresponding to $[v,l^\prime]$ such that $DFSState(l^\prime)=$``childBridge'' is connected to $[v^\prime,
l^{\prime\prime}]$, where $v^\prime$ is the child of $v$ in the
DFS-tree through the port $l^\prime$, and $l^{\prime\prime}$ is the
port of $v^\prime$ corresponding to $l^\prime$.
\item The ordered bag corresponding to $[v,l^\prime]$ such that $DFSState(l^\prime)=$``parentBridge'' is connected to $[v^\prime, l^{\prime\prime}]$, where $v^\prime$ is
the parent of $v$ in the BFS-tree through the port $l^\prime$, and
$l^{\prime\prime}$ is the port of $v^\prime$ corresponding to
$l^\prime$.
\end{itemize}
\item Starting from the requesting node, connect together the ordered tree decompositions of the blocks by using the virtual
nodes to construct a complete ordered tree decomposition of rank $2$
for the whole network.
\end{enumerate}

\section{Distributed model-checking of MSO over networks with bounded degree and bounded tree-length}\label{appendix-tree-decomp-btl}

We illustrate the proof of Theorem~\ref{thm:MSO-BTL-frugal} by
considering the asynchronous distributed systems with a BFS tree
pre-computed and stored on nodes of the network.

The distributed algorithm includes the following four phases.
\vspace*{-2mm}
\begin{description}
\item[Phase I]: At first, we show that an ordered tree decomposition $\mathcal{T}$ with rank $f(d,t)$ (for some function $f$) and with width at most $g(d,t)$ (for some function $g$) can be
distributively constructed within the complexity bounds.
\item[Phase II]: Then from the ordered tree decomposition, a labeled tree $\mathcal{T}^\prime$
over some finite alphabet $\Sigma$ with rank $f(d,t)$ can be
obtained easily.
\item[Phase III]: From an MSO sentence $\varphi$, a deterministic bottom-up automaton
$\mathcal{A}_{\varphi}$ over $f(d,t)$-ary $\Sigma$-labeled trees can
be constructed.
\item[Phase IV]: At last we show that $\mathcal{A}_{\varphi}$ can be distributively run over
$\mathcal{T}^\prime$ within the complexity bounds.
\end{description}
\vspace*{-2mm}



\noindent For the proof of \textbf{Phase I}, we introduce the notion
of BFS-layering tree used in \cite{DG04}.

Let $s$ be a distinguished vertex in graph $G=(V,E)$. Let $L^i=\{v
\in V | dist_G(s,v) = i\}$. A \emph{layering partition} of $G$ is a
partition of each set $L^i$ into $L^i_1,\cdots, L^i_{p_i}$ such that
$u,v \in L^i_j$ if and only if there exists a path from $u$ to $v$
such that all the intermediate vertices $w$ on the path satisfy that
$dist_G(s,w) \ge i$.

Let $H=(V_H,E_H)$ be the graph defined as follows:
\begin{itemize}
\item $V_H$: the sets $L^i_j$.

\item $E_H$: $\{L^i_j,L^{i^\prime}_{j^\prime}\} \in E_H$ if and only
if there is $u \in L^i_j$ and $v \in L^{i^\prime}_{j^\prime}$ such
that $(u,v) \in E$.
\end{itemize}

\vspace*{-2mm}
\begin{theorem}
\cite{DG04} The graph $H$ is a tree.
\end{theorem}
\vspace*{-2mm}

$H$ is called the \emph{BFS-layering tree} of $G$.

$H$ can be seen as a rooted labeled tree $(I,F,r,B)$ with
\begin{itemize}
\item $I=\{(i,j)| L^i_j \in V_H\}$, $F=\{\{(i,j),(i^\prime,j^\prime)\} |
\{L^i_j,L^{i^\prime}_{j^\prime}\} \in E_H\}$,

\item The root $r$ is $(0,1)$,

\item $B((i,j))=L^i_j$.
\end{itemize}

If we replace the label of $(i,j)$ by $B(i,j) \cup
B(i^\prime,j^\prime)$ in $H$, where $(i^\prime,j^\prime)$ is the
parent of $(i,j)$ in $H$, then we get a new rooted labeled tree
$\mathcal{S}=(I,F,r,L)$.

\vspace*{-2mm}
\begin{theorem}\label{thm:BFS-layering-tree-decomp-2}
\cite{DG04} $\mathcal{S}$ is a tree decomposition of $G$ such that
$length(\mathcal{S}) \le 3 \cdot tl(G)+1$.
\end{theorem}
\vspace*{-2mm}

$\mathcal{S}$ is called the \emph{BFS-layering tree decomposition}
of $G$.

\vspace*{-2mm}
\begin{lemma}\label{lem:bounded-degree-TL-TW}
If a graph $G$ has bounded degree $d$ and bounded tree-length $t$,
then the width and the rank of the BFS-layering tree decomposition
$\mathcal{S}$ are bounded by $f(d,t)$ and $g(d,t)$ respectively for
some functions $f$ and $g$.
\end{lemma}
\vspace*{-2mm}

\vspace*{-2mm}
\begin{proof}
The fact that there is a function $f$ such that the width of
$\mathcal{S}$ is bounded by $f(d,t)$ follows directly from the
bounded degree and bounded tree-length assumption, and
Theorem~\ref{thm:BFS-layering-tree-decomp-2}.

The rank of the BFS-layering tree decomposition $\mathcal{S}$ is the
rank of the BFS-layering tree $H$.

Since the length of each bag of the BFS-layering tree decomposition
$\mathcal{S}$ is bounded by $3t+1$, the size of each layering
partition $L^i_j$ of $H$ is bounded by
$1+d+\cdots+d^{3t+1}=\left(d^{3t+2}-1\right)/(d-1)$. Each node in
BFS-layer $i$ has at most $(d-1)$-neighbors in BFS-layers greater
than $i$. So the number of children of node $(i,j)$ in $H$ is no
more than $(d-1)\cdot\left(d^{3t+2}-1\right)/(d-1)=d^{3t+2}-1$. Let
$g(d,t)=d^{3t+2}-1$. \qed
\end{proof}
\vspace*{-2mm}


\noindent In the following, we design a distributed algorithm to
construct $\mathcal{S}$, then an ordered tree decomposition
$\mathcal{T}$ from $\mathcal{S}$ to finish \textbf{Phase I}. The
distributed algorithm consists of two stages:
\begin{description}
\item[Stage 1]: Construct the BFS-layering tree $H$
bottom-up. Because $G$ has bounded tree-length $t$, $\mathcal{S}$
has length no more than $3t+1$ by
Theorem~\ref{thm:BFS-layering-tree-decomp-2}. Thus if two nodes of
layer $L^i$ are in the same layering partition, then the distance
between them is no more than $3t+1$. Consequently, when the layering
partition of $L^{i+1}$ has been computed, each node $v$ in $L^i$ can
know which nodes are in the same layering partition of $L^i$ by
doing some local computation in its $(3t+1)$-neighborhood.
%
\item [Stage 2]: Construct the BFS-layering tree decomposition $\mathcal{S}$ from $H$, and an ordered tree decomposition $\mathcal{T}$ from $\mathcal{S}$.
\end{description}

\noindent Suppose each node $v$ stores its unique identifer $ID$,
its depth in the pre-computed BFS-tree, $depth(v)$, and the depth of
the BFS-tree, $treeDepth$.
\smallskip

\noindent \textbf{Stage 1: BFS layering tree construction}

The requesting node sends messages to ask each node $v$ to classify
its ports with state ``non-tree'' into ports with state ``upward'',
``downward'' or ``horizon'' as follows: For each port $l$ of $v$
with state ``non-tree'', let $w$ be the node connected to $v$
through $l$, if $depth(w)< depth(v)$, then $state(l):=$``upward'';
if $depth(w) > depth(v)$, then $state(l):=$``downward''; if
$depth(w) = depth(v)$, $state(l):=$``horizon''.

Then the requesting node $v_0$ sends message
STARTLAYERING$(treeDepth)$ over all ports.

For each node $v$, when it receives message STARTLAYERING$(td)$, it
does the following:

If $v$ is not a leaf, it sends STARTLAYERING$(td)$ over all ports
$l^\prime$ such that $state(l^\prime)$=``child''.

Otherwise, if $depth(v)=td$, $v$ starts the BFS-Layering tree
construction as follows:
\begin{itemize}
\item Node $v$ collects the information in its $(3t
+1)$-neighborhood, and determines the set of nodes that are in the
same layering partition of $L^{depth(v)}$ as $v$, denoted by
$P^{depth(v)}(v)$.

\item Then $v$ sets $partitOver(v):=true$ and sets $partitId:= \min\{w.ID | w \in
P^{depth(v)}(v)\}$.

\item At last $v$ sends LAYERPARTIT$(partitId)$ over
all ports with state ``parent'' or ``upward''.

\end{itemize}

For each node $v$, when it receives message
LAYERPARTIT$(downPartId)$ over the port $l$, it does the
following:\\

\noindent
  \begin{tabular*}{\textwidth}{l@{\extracolsep{\fill}}}
    \hline
    Message LAYERPARTIT$(downPartitId)$ over the port $l$\\
    \hline
    1: $downPartitOver(l):=true$.\\
    2: $downPartitId(l):=downPartitId$.\\
    3: \textbf{if} each port $l^\prime$ with state ``child'' or ``downward'' satisfies $downPartitOver(l^\prime)=true$ \textbf{then} \\
    4: \hspace*{1em} \textbf{if} $v$ is not the requesting node \textbf{then}\\
    5: \hspace*{1em}\hspace*{1em} Collects all the information in $N_{3t +1}(v)$.\\
    6: \hspace*{1em}\hspace*{1em} \textbf{if} each node $w$ in $N_{3t+1}(v)$
    such that $depth(w)=depth(v)$ satisfies the condition: each port $l^\prime$ of $w$\\
    \hspace*{1em}\hspace*{1em}\hspace*{1em}\hspace*{1em}  such
    that $state(l^\prime)=$``child'' or ``downward'' satisfies $downPartitOver(l^\prime)=true$
    \textbf{then}\\
    7: \hspace*{1em}\hspace*{1em}\hspace*{1em} $P^{depth(v)}(v):=$ the set of nodes in the same layering partition of $L^{depth(v)}$ as
    $v$.\\
    8: \hspace*{1em}\hspace*{1em}\hspace*{1em} $partitId(v):=\min\{ID(w)|w \in P^{depth(v)}(v)\}$.\\
    9: \hspace*{1em}\hspace*{1em}\hspace*{1em} $partitOver(v):=true$.\\
    10:\hspace*{1em}\hspace*{1em}\hspace*{1em} \textbf{if} $ID(v)=partitId(v)$ \textbf{then}\\
    11:\hspace*{1em}\hspace*{1em}\hspace*{1em}\hspace*{1em} $partitIdList(v):=$ a list enumerating the elements in $P^{depth(v)}(v)$.\\
    12:\hspace*{1em}\hspace*{1em}\hspace*{1em} \textbf{end if}\\
    13:\hspace*{1em}\hspace*{1em}\hspace*{1em} \textbf{for} each $w \in N_{3t+1}(v) \backslash
    P^{depth(v)}(v)$ such that $depth(w)=depth(v)$ and $partitOver(w)=false$
    \textbf{do}\\
    14:\hspace*{1em}\hspace*{1em}\hspace*{1em}\hspace*{1em} $v$ sends
    message LAYERPARTITNOW$(ID(v), ID(w), 3t+1)$ over all ports.\\
    15:\hspace*{1em}\hspace*{1em}\hspace*{1em} \textbf{end for}\\
    16:\hspace*{1em}\hspace*{1em}\hspace*{1em} \textbf{for} each
    vertex $w \in P^{depth(v)}(v)$, $w \neq v$ \textbf{do}\\
    17:\hspace*{1em}\hspace*{1em}\hspace*{1em}\hspace*{1em} $v$ sends message PARTITID$(ID(w), partitId(v),3t+1)$ over all ports.\\
    18:\hspace*{1em}\hspace*{1em}\hspace*{1em} \textbf{end for}\\
    19:\hspace*{1em}\hspace*{1em}\hspace*{1em} $v$ sends LAYERPARTIT$(partitId(v))$ over all ports with state
    ``parent'' or ``upward''.\\
    20:\hspace*{1em}\hspace*{1em} \textbf{end if}\\
    21:\hspace*{1em} \textbf{end if}\\
    22: \textbf{end if}\\
    \hline
  \end{tabular*}
\\

When node $v$ receives message PARTITID$(dest, id, hop)$ over the
port $l$, it does the following:
\begin{itemize}
\item If $ID(v)=dest$: it does the following:
\begin{itemize}
\item If $partitOver(v)=false$ then it sets \\
$partitOver(v):=true$, $partitId(v):=id$, and sends
LAYERPARTIT$(partitId(v))$ over all ports with state ``parent'' or
``upward''.
\item Otherwise: it does nothing.
\end{itemize}
\item Otherwise, if $hop>0$, $v$ sends
PARTITID$(dest, id, hop-1)$ over all the ports except $l$ unless it
has already done so.
\end{itemize}

When node $v$ receives message LAYERPARTITNOW$(src, dest, hop)$ over
the port $l$, it does the following:
\begin{itemize}
\item If $ID(v)=dest$: if $partitOver(v)=false$, then $v$ executes Lines 5-20 of the
pseudo-code above.

\item Otherwise: if $hop>0$, $v$ sends LAYERPARTITNOW$(src, dest,
hop-1)$ over all the ports except $l$ unless it has already received
some message LAYERPARTITNOW with $src$ as the source.
\end{itemize}

\noindent \textit{Complexity analysis of }\textbf{Stage 1}:

During the distributed computation of \textbf{Stage 1}, each node
$v$ receives at most $k$ LAYERPARTIT messages from its children.
Moreover, $v$ only receives messages LAYERPARTITNOW with nodes in
$N_{3t+1}(v)$ as the source vertex. Node $v$ starts the local
computation to collect the information in $N_{3t+1}(v)$ only when
receiving messages LAYERPARTIT or LAYERPARTITNOW, so $v$ only starts
$O(1)$ local computations. Since each node only participates in the
local computations started by the nodes in $N_{3t+1}(v)$, $v$ only
participates in $O(1)$ local computations, it follows that only
$O(1)$ messages are sent over each link during the distributed
computation of \textbf{Stage 1}.

\medskip

\noindent \textbf{Stage 2: Ordered BFS-layering tree decomposition
construction}

When the requesting node receives messages
LAYERPARTIT$(downPartitId)$ from all its ports, it sends message
LAYERTREEDECOMP over all the ports.

Each node $v$ receiving message LAYERTREEDECOMP from the port $l$
sends LAYERTREEDECOMP over all its ports with state ``child''. If
$ID(v)=partitId(v)$, it sends message GETPARENTPARTIT over the port
with state ``parent''.

When node $v$ receives message GETPARENTPARTIT from the port $l$,
$v$ sends messages to get $partitIdList(w)$ from the node $w$ such
that $ID(w)=partitId(v)$. Then it sends
RETURNPARENTPARTIT$(partitIdList(w))$ over the port $l$.

When a node $v$ receives RETURNPARENTPARTIT$(idList)$ from the port
$l$,
\begin{itemize}
\item it sets $bagIdList(v):=list_{f(d,t)}(partitIdList(v) \cdot idList),$

where $list_{f(d,t)}(p)$ generates a list of length $f(d,t)$ from
the list $p$ (of length less than $f(d,t)$) by repeating the last
element in $p$;

\item if $v$ is a leaf or $v$ has received messages DECOMPOVER
from all its ports with state ``child'', then $v$ sends message
DECOMPOVER over the port $l^\prime$ with state ``parent''.
\end{itemize}

When a node $v$ receives DECOMPOVER from the port $l$, if it has
received messages DECOMPOVER from all its ports with state
``child'', then $v$ sends message DECOMPOVER over the port
$l^\prime$ with state ``parent''.

\bigskip

\noindent Now we consider \textbf{Phase II-IV} of the proof of
Theorem~\ref{thm:MSO-BTL-frugal}.

When the requesting node receives messages DECOMPOVER from all its
ports, the computation of \textbf{Phase I} is over.

\noindent \textbf{Phase III} is done by the following lemma.

\begin{lemma}\label{thm:MSO-automata}
\cite{FlumG06} Given an MSO sentence $\varphi$ and $d, t \in
\mathds{N}$, let $f$ and $g$ be the functions in
Lemma~\ref{lem:bounded-degree-TL-TW}, then a deterministic bottom-up
$g(d,t)$-ary tree automaton $\mathcal{A}$ over alphabet
$\Sigma_{f(d,t)}$ can be constructed from $\varphi$ such that:

For each graph $G$ with degree bounded by $d$ and tree-length at
most $t$, and each ordered tree decomposition $\mathcal{T}=(I,F,r,
L)$ of $G$ with width $f(d,t)$ and rank at most $g(d,t)$, we have
that $G \models \varphi$ if and only if $\mathcal{A}$ accepts
$\mathcal{T}^\prime$, where $\mathcal{T}^\prime$ is the rooted
$\Sigma_{f(d,t)}$-labeled tree obtained from $\mathcal{T}$.
\end{lemma}

\noindent Now we consider \textbf{Phase II} and \textbf{Phase IV}.

When the requesting node receives messages DECOMPOVER from all its
ports, then it knows that the construction of the ordered
BFS-layering tree decomposition $\mathcal{T}$ has been done.

Then the requesting node constructs automaton $\mathcal{A}$ from
$\varphi$ and broadcasts it to all the nodes in the network.

Afterwards, the requesting node starts the computation to relabel
$\mathcal{T}$ and get the rooted labeled tree $\mathcal{T}^\prime$
in a way similar to the \textbf{Stage 2} in \textbf{Phase I}.

Finally automaton $\mathcal{A}$ can be run over $\mathcal{T}^\prime$
in a bottom-up way similar to the \textbf{Stage 1} in \textbf{Phase
I}.

The proof of Theorem~\ref{thm:MSO-BTL-frugal} is completed. \qed

\end{appendix}
\end{document}